\documentclass[fleqn,usenatbib]{mnras}
\usepackage{newtxtext,newtxmath}
\usepackage{newtxtext,newtxmath}
\usepackage{hyperref}
\usepackage[T1]{fontenc}
\usepackage{ae,aecompl}

\usepackage{graphicx}	
\usepackage{amsmath}	
\usepackage{amsfonts}	
\usepackage{hyperref}
\usepackage{natbib}
\usepackage[caption=false]{subfig}
\usepackage{ragged2e}
\usepackage{placeins}
\usepackage{bigints} 
\usepackage{physics}
\usepackage{xcolor}
\usepackage{cleveref}
\usepackage[export]{adjustbox}
\usepackage{ulem}
\usepackage{bm}

\title[Sound waves in the ICM]{Acoustic waves and g-mode turbulence as energy carriers in a viscous intracluster medium}

\author[]
{Prakriti Pal Choudhury \thanks{Email: pp512@cam.ac.uk}, Christopher S. Reynolds\\
$^\ddag$ Institute of Astronomy, University of Cambridge,Madingley Rd, Cambridge CB3 0HA, United Kingdom}

\pubyear{2022}

\begin{document}
\label{firstpage}
\pagerange{\pageref{firstpage}--\pageref{lastpage}}
\maketitle
\begin{abstract}
Many recent works on the observed galaxy clusters in the X-rays highlight broadly two classes of exclusive energy carriers - sound waves and turbulence. In order to understand this dichotomy, we design an idealized three-dimensional hydrodynamic simulation of a cluster, to assess which of these carriers can dissipate energy in and around the core ($\gtrsim 100$ kpc) . Specifically, we explore how gentle (long-duration outbursts) and intermediate (shorter duration outbursts) feedback modes can function efficiently mediated by compressible (sound waves) and incompressible (g-modes/instabilities/turbulence) disturbances. Since g-modes are confined tightly to the central core, we attempt to maximise the flux of fast sound waves to distribute the feedback energy over a large distance. We find that the contribution to heat dissipation from sound and turbulence varies on the basis of the aforementioned feedback modes, namely: turbulence contributes relatively more than sound in the slow-piston regime and vice versa for the intermediate regime. For the first time in a 3D simulation, we show that up to $\lesssim 20\%$ (in some directions) of the injected power can be carried away by sound flux in the intermediate feedback but it reduces to $\lesssim 10 \%$ (in some directions) in the slow-piston regime. Lastly, we find that sound waves can be elusive if we deduce the equation-of-state (isobaric/isentropic) of the fluctuations from X-ray observations.

\end{abstract}

\begin{keywords}
galaxies: clusters: intracluster medium -- hydrodynamics 
\end{keywords}



\section{Introduction}
\label{sec:intro}
The intracluster medium (ICM) is a host to diverse energetic cosmic events. One of the most important among these events is the regular ejection of large amounts of energy in and around the central galaxy. This happens due to accretion by the central supermassive black hole (SMBH) which enables not just the infalling gas to release the gravitational energy but also triggers high-speed jets powered by magnetic processes and which terminate in large-scale ($\sim {\rm kpc}$) radio lobes. Irrespective of the underlying mechanism, the release of energy in any form (kinetic and/or thermal and/or radiative) is collectively called feedback since it helps to prevent excessive cooling (``cooling flows''), quench star formation, and reduces accretion rate (\citealt{1994ARA&A_fabian, 2012ARA&A_fabian, 2002ApJ_cavaliere, 2004cbhg.symp_begelman, 2005ApJ_murray, DiMatteo2005, 2007MNRAS_sijacki} and references therein). In the absence of feedback, both star formation and central galaxy growth are expected to happen faster leading to a catastrophically cooling core (inner $\sim 100$ kpc) which is not seen in observations. Instead the ICM is seen in X-rays as a stable hot, diffuse medium close to a global thermal equilibrium.

The quantitative estimate of heating required to offset excessive cooling is easily accounted for in the work done to inflate radio cavities by the jets (\citealt{2000MNRAS_fabian, 2004ApJ_birzan}). Two decades of intensive exploration in observations and hydrodynamic simulations based on feedback regulated cooling in clusters/galaxies lead to modeling radio jets tied up with accretion rates into the SMBH (\citealt{2006MNRAS_allen, 2006ApJ_rafferty, 2006MNRAS_best, 2007MNRAS_merloni, 2007ApJ_wise, 2007annurev_mcnamara, 2012MNRAS_dubois, 2014ApJ_li, 2015ApJ_prasad}). A broad discussion on mechanical feedback via jets and related unresolved problems have been presented by \citealt{McNamara_2012}. Feedback from quasars in the form of winds and radiation has also been discussed in the context of ICM and massive galaxies for a range of redshifts (\citealt{1990MNRAS_fabian, 2008A&A_nesvadba, 2011ApJ_greene}).

Despite a strong consensus on the role of feedback, particularly on the adequacy of energy from high-speed bipolar jets powered by microphysical magnetized accretion process, a major concern has been how to isotropize the available energy (\citealt{2006ApJ_vernaleo}). Strong shocks driven by the terminal working surface of powerful jets can heat the gas locally, but the ICM away from the jet axis will remain unheated. Precessing jets can solve this problem partially by generating radio bubbles in multiple directions (\citealt{2006MNRAS_dunn, 2008MNRAS_sternberg}). But localization of the equivalent energy (enthalpy associated with the bubbles and the thermal energy deposited at bubble-ICM interfaces) is still an issue. Extremely powerful jets can prevent cooling flows by overheating and disrupting the core intermittently which is commonly seen in cosmological simulations. However, many observed clusters are not as disturbed as is expected from powerful events (e.g., Perseus, see \citealt{2006MNRAS_fabian}). This motivates the search for an efficient mechanism to distribute energy from moderately powered feedback. Note that even in quasar mode feedback (via winds/radiation), it is expected that gas is thermalized closer to SMBH and the localization problem persists.

The problem of energy distribution in a simplified form reduces to the following: there is sufficient thermal energy available nearer to the core or away from center along any collimated jet. Thus we need mechanisms for homogenous and isotropic distribution of this energy in and around the core. The obvious mode of transport are sound waves which are fast and ubiquitous (but also see \citealt{2007ApJ_fujita} for a discussion on heating by Alfv\'en waves). Perseus, which is the most well studied galaxy cluster, hosts quasi-concentric ripples in X-ray surface brightness interpreted as sound flux (\citealt{2007MNRAS_sanders}). The flux diminishes beyond $40-50$ kpc possibly due to sound dissipation. This mode of transport assures that there is energy circulation without a flow of matter and hence it is a gentle feedback.

Theoretical discussions on acoustic energy propagation (e.g., \citealt{2018ApJ_zweibel}) reveal that plasma transport processes (viscous+conductive) must be reduced for sound to not dissipate too close to the source. A good argument for suppression of transport is valid for galaxy clusters since magnetic fields are present which are known to suppress the perpendicular (cross-field) transport to a great extent (\citealt{2012MNRAS_parrish, 2016ApJ_roberg-clark}). Hence it is expected that long wavelength sound waves may actually escape to the outer core and satisfy the requirement of a long inertial range before damping can affect. For more than a decade, sound waves/ripples have been explored in more observed clusters (e.g., \citealt{2008MNRAS_sanders}) and hydrodynamic simulations to understand the feasibility of the heating mechanism and escape fraction to higher atmospheres in the ICM (\citealt{2004ApJ_ruszkowski, 2009MNRAS_shabala, 2009MNRAS_sternberg, 2018MNRAS_tang, 2019ApJ_bambic}). While the heating by waves has not been explored sufficiently, the escape fraction has been found to be between $\sim 12$\% (\citealt{2017MNRAS_tang}) in simple 1D ICM models to $\gtrsim 25$\% in stratified 2D intracluster medium with powerful jets (\citealt{2019ApJ_bambic}). Most recently, \citealt{2022arXiv_wang} claimed that the contribution of sound wave to heating is negligible compared to strong shocks for very powerful feedback. 

Another strong evidence of gentle distribution of energy comes from measurement of the gas velocities or turbulence using soft X-ray spectra from {\tt Hitomi} observation (\citealt{2016Natur_hitomi}). Within the central $60$ kpc of Perseus, it is inferred from the {\tt Hitomi} data that the turbulent pressure support is at most $4\%$ of thermal pressure and hence insignificant. It is possible that the turbulent motions are dissipated fast within the core and a fresh episode of stirring refills the core with moderate gas motions intermittently. Analysis of X-ray surface brightness fluctuations in Perseus/M87/Virgo (e.g., \citealt{2016MNRAS_zhuravleva, 2018ApJ_zhuravleva, 2016ApJ_arevalo}) distinguish between bubble enthalpy and energy carried away by shocks (or waves), and claim that the former dominates the AGN energy budget promoting turbulent motions. Possibly such bubble induced motions have been captured by {\tt Hitomi}. However, as \citealt{2017MNRAS_fabian} discuss, the problem of radial transport of feedback energy is not solved until a volume-filling turbulence can be triggered with the constraint of moderate velocity dispersion imposed by {\tt Hitomi}. \citealt{2018MNRAS_zhang} discuss how g-modes triggered by AGN bubbles may have terminal velocities within the constraints; but also see \citealt{2015ApJ_reynolds} for a discussion on the inefficiency of driving volume-filling g-mode turbulence. Note that high-$\beta$ plasma like ICM may have Alfv\'enic turbulence as well e.g., \citealt{PhysRevLett_squire} or other instabilities e.g., \citealt{2016MNRAS_melville} which may contribute to volume-filling turbulence. It is not well established if such turbulence can heat the gas enough in and around the cores.

In this work, we discuss the energy carriers in a hydrodynamic viscous and radiatively cooling intracluster medium embedded in a dark matter halo potential. We explore how a moderately powered repeated central injection of thermal energy with a given duty cycle can generate both turbulence and sound flux. The goal is to maximise the production of sound wave trains hence we inject only pure thermal feedback. Taking a cue from Perseus on gentle flow of energy across the core, we explore slow to moderate rates of energy injection to model weak shocks ($t_{\rm b}/t_{\rm E}=[0.5,2]$; see \citealt{2017MNRAS_tang} and section \ref{sec:ws_lit}). We assess the contributions of the two carriers in triggering gas motions and subsequent viscous damping of the motions leading to thermalization. In section \ref{sec: physset} we include complete details of the physical modeling, motivation and description of the physical set-up, simulation parameters and diagnostics. We also provide a basic linear theory of sound waves and g-modes in a viscous ICM. In section \ref{sec: res} we present the results from our suite of simulations. In section \ref{sec: disc} we discuss the implications of the results and in section \ref{sec:cons} we include our conclusions.

\section{Physical model and numerical set-up}
\label{sec: physset}
In this section we describe the model to study the circulation of energy in a viscous ICM using 3D hydrodynamic simulations. This model, while being idealised, is very useful to understand how energy from a gentle feedback can be distributed across the bulk ICM.

\subsection{Basic equations}
The basic equations for compressible ideal hydrodynamics that are used to study the intracluster medium are:
\begin{eqnarray}
\label{eq:eq1}
\frac{D \rho}{Dt} &=& - \rho {\mathbf \nabla} \cdot {\mathbf{u}} \\
\label{eq:eq2}
\frac{D {\mathbf{u}}}{Dt} &=& -\frac{1}{\rho} {\mathbf \nabla} p + \frac{1}{\rho}  {\mathbf \nabla}\cdot \Pi - g\left(r\right) {\hat{r}} \\
\label{eq:eq3}
\frac{p}{(\gamma-1)} \frac{D}{Dt} \left [ \ln \left ( \frac{p}{\rho^\gamma} \right ) \right ] &=& - q^-(\rho, T) + q^+_{\rm ws} + {\mathbf \nabla}\cdot ({\mathbf{u}}\cdot \Pi)
\end{eqnarray} 
where $\frac{D}{Dt}$ is the Lagrangian derivative, $\rho$, $\bm{v}$ and $p$ are mass density, velocity and pressure; $\gamma=\frac{5}{3}$ is the adiabatic index; $g\left(r\right)$ is the acceleration due to gravity. The viscous stress tensor is denoted by $\Pi_{ij} = \eta [(\partial {\bm u}_i/ \partial x_j + \partial {\bm u}_j/ \partial x_i) - 2 {\bf \nabla} \cdot \mathbf{ u} \delta_{ij} /3]$, $\eta = \rho \nu (\nu \propto T^{5/2})$ is the shear viscosity, $q^-$ is radiative cooling rate and $q^+_{\rm ws}$ is the weak-shock energy injection rate. 

We use the {\tt PLUTO} code (\citealt{2007ApJS_mignone}) to evolve the hydrodynamic equations. The code evolves the equations in conservative form with static gravity, viscosity (supertimestepping), and an ideal equation of state. The equations are solved using the {\tt HLL} Riemann solver (a few simulations have been carried out using {\tt HLLC}; see Table \ref{tab:sims}), linear reconstruction and second-order Runge-Kutta timestepping. All the parameters and initial conditions specific to the problem are described in section \ref{sec: setup} and section \ref{sec: coolht}.

\subsection{Simulation set-up}
\label{sec: setup}
The initial temperature profile is based on the empirical fit to surface brightness profiles as given in eq. 7 (and the references in section 2.2) in \citealt{2016ApJ_yang}. 
\begin{eqnarray}
\label{eq:initemp}
T = T_0 \frac{1 + {(r/r_0)}^3}{2.3 + {(r/r_0)}^3} {[1 + {(r/r_1)}^2]}^{-0.32}
\end{eqnarray}
where $T_0 = 7$ keV, $r_0=71$ kpc, and $r_1=380$ kpc. Using this form we analytically solve the hydrostatic equilibrium in NFW gravity to obtain radial profiles of thermal pressure. 
\begin{eqnarray}
\frac{dp(r)}{dr} &=& -\rho (r) g_{\rm NFW}(r) \\
\nonumber
g_{\rm NFW}(r) &=& \frac{GM_{\rm halo}}{\ln(1+c) - c/(1+c)}\Big( \frac{\ln(1+r/r_{\rm s})}{r^2} \\
&-& \frac{1}{r_{\rm s} r (1 + r/r_{\rm s})} \Big)
\end{eqnarray}
where $M_{\rm halo} = 5 \times 10^{14}~M_{\odot}$, $c = 6.81$, and $r_{\rm s} = r_{\rm halo}/c$. Here $r_{\rm halo}$ is defined to be the virial radius within which the mean density is $200$ times the critical density of the Universe.
We fit a profile to the pressure as a function of radius and initialize that in a cartesian 3D grid in {\tt PLUTO}. Using the fitted pressure profile and empirical temperature, we calculate the density in the initialization. The pressure profiles are fitted as sum of decaying exponentials. 
\begin{eqnarray}
\nonumber
p_0 \Big( A e^{-a r} + B e^{-b r} + (1.0 - A - B) e^{-c r} \Big)
\end{eqnarray}
where $r = (x^2 +y^2 + z^2)^{\frac{1}{2}}$ and $A, B, a, b, c$ are fitting parameters and $p_0$ is the maximum presure at the center for our solution of hydrostatic equilibrium. These profiles are extrapolated to the boundary zone and velocities are always fixed to zero in the boundary to minimise any reflection or free-flow into the domain from boundary zones. There is an initial isobaric density perturbation field embedded on the equilibrium density profile in the box to break the spherical symmetry, such that any random mode amplitude is $\propto k^{-1}$ ($k^2 = k^2_{\rm x} + k^2_{\rm y} + k^2_{\rm z}$). These modes smooth out fast in our simulations and do not affect the dynamics after a short time. We carry out a few test runs removing the initial modes above $k > k_{\rm max}$ (where the equivalent length scale is $\lesssim 10$ kpc) and do not see significant difference in results. The decay of such perturbations are fast enough to not even produce small-scale thermal instability. We typically see some evidence of condensation in larger length scales along a rough spiral moving inward (black regions in lower right panel of Figure \ref{fig:obs0}). 

The kinematic viscosity for a fully ionized hydrogen plasma is given by,
\begin{eqnarray}
\nu = 1.\times 10^{25} {T_7}^{\frac{5}{2}} n^{-1} {\xi}_{\nu}\ \ {\rm cm}^2\,{\rm s}^{-1}
\end{eqnarray}
where $T_7$ is in the units of $10^7$ K and ${\xi}_{\nu}$ is the fraction by which viscosity is suppressed from the unmagnetized values. The above equation is same as eq. 3 in \citealt{2005MNRAS_fabian}. We adopt a fiducial ${\xi}_{\nu} = 0.1$ motivated by the best fit parameters suggested by \citealt{2005MNRAS_fabian} (see Fig 1 and section 2 for a discussion on the relatively greater importance of viscous over conductive heating for linearly propagating sound waves). 

All the simulations have equal number of points in the three directions, $N_x = N_y = N_z$, in a cartesian cubic (${400~{\rm kpc}}^3$) box with the ICM centered in it. We have an inner uniform cube which encloses $100$ kpc ($\times 2$ for diameter) sphere and $3$ pairs of half as resolved $100$ kpc regions on either sides in each direction surrounding the inner cube. Details for the set of simulations discussed in this paper are tabulated in Table \ref{tab:sims}.

\subsection{Radiative cooling and energy injection to generate weak shocks}
\label{sec: coolht}
The gas undergoes radiative cooling in sub-cycles between successive hydrodynamic time steps. We use the exact integration scheme described in \citealt{2009ApJS_townsend}. The cooling is shut off below $10^4$\,K. The weak-shock feedback energy (described below) is also injected in same sub-cycles. The radiative cooling rate is expressed as:
\begin{eqnarray}
q^{-} = n_e n_i \Lambda (T)
\end{eqnarray} 
where $n_e, n_i$ are number densities of electrons and ions and $\Lambda$ denotes a cooling function. We use a tabulated cooling function based on \citealt{1993ApJS_suthdop}.  

In order to produce weak shocks, we inject thermal energy $E_{\rm inj}$ (see sections \ref{sec:ws_lit}, \ref{sec:desc_tab} and Table \ref{tab:sims}) over a time period of $t_{\rm spread}$. The energy is injected within a radius of $R_{\rm inj}$ uniformly. The injection is repeated every $100$ Myr (fiducial) for the next $t=t_{\rm spread}$ and shut off until the next episode. The rate of energy injection in each cell of the simulation is:
\begin{eqnarray}
q^+_{\rm ws} = \frac{E_{\rm inj}}{\frac{4}{3}\pi R^3_{\rm inj} t_{\rm spread}}
\end{eqnarray}
This implies that at any point of time within the duration of a single outburst, the total power injected is $E_{\rm inj}/t_{\rm spread}$ ergs/s. While the $E_{\rm inj}$ in our simulations is in the range of estimated net energy injection in typical observed clusters, depending on the duration $t_{\rm spread}$ we model gentle to moderately strong feedback (see section \ref{sec:ws_lit} for previous literature and section \ref{sec:desc_tab} for a description of the parameters we use in the suite of simulations).

\subsubsection{How do these outbursts compare with the models of weak-shocks?}
\label{sec:ws_lit}
Following \citealt{2017MNRAS_tang}, there are three important phases in the lifetime of an outburst, namely, free-expansion: when the swept-up mass is less than the injected mass and its thermal energy is $\ll E_{\rm inj}$,    Sedov-Taylor (ST) phase (\citealt{1946RSPSA_taylor, 1959sdmm.book_sedov}): when the swept-up mass is significant but its thermal energy is $\lesssim E_{\rm inj}$, and lastly, the wave-like phase: when the thermal energy of the swept-up material is $>E_{\rm inj}$. For AGN feedback, it is expected that the mass injected will be always negligible and the transition from ST to wave-like is the most interesting to interpret the observed weak-shocks in clusters. This is also the appropriate regime to determine the role of sound waves in AGN feedback. A rough estimate of the spatial and temporal location of such a transition to wave-like regime is given by the radius at which the thermal energy of the swept-up gas is equal to $E_{\rm inj}$,
\begin{eqnarray}
\nonumber
4 \pi R^3_{\rm E} P_{\rm ambient}/3 = E_{\rm inj}
\end{eqnarray} 
where $P_{\rm ambient}$ varies very slowly across a few kpc and hence $R_E$ calculated as above is approximately same when calculated taking into account the decrease in $P_{\rm ambient}$ with radius (we verified this). The ambient pressure in Perseus, M87 (as mentioned in Table 1 in \citealt{2017MNRAS_tang}) and this model are $0.35~{\rm keV {cm}^{-3}}$, $7.8 \times {10}^{-2}~{\rm keV {cm}^{-3}}$, $0.21~{\rm keV {cm}^{-3}}$ respectively. In our set-up, $R_{\rm E}= 6.27$ kpc and $12.43$ kpc (fiducial). Using Table 2 of \citealt{2017MNRAS_tang}, we can estimate $R_{\rm E}$ for Perseus ($E_{\rm inj, Perseus} = 9.67\times {10}^{58}$ ergs) and M87 ($E_{\rm inj, M87} = 5.03\times {10}^{57}$ ergs) to be $11.19$ kpc and $6.89$ kpc respectively. The ejecta radius estimated for these clusters are $\sim 0.6 R_{\rm E}$ which is smaller than our $R_{\rm inj}$ (fiducial) in our simulation (but note we do a low resolution simulation, $L_{\rm 0.66 R_E}$, in Table \ref{tab:sims}). Also, note that the ejecta radius is expected to increase within $t<t_{\rm spread}$ during any single outburst. 

The sound-crossing time across $R_{\rm E}$, $t_{\rm E} (= R_{\rm E}/c_{\rm s, ambient})$, is a characteristic timescale that determines the impact of the outburst in terms of energy partition according to the theory and simulations proposed by \citealt{2017MNRAS_tang}. A significantly large strongly-shocked (SS) envelope is expected in the ``instantaneous outburst" regime ($t_{\rm spread} \ll t_{\rm E}$; note in \citealt{2017MNRAS_tang} the same parameter is noted as $t_{\rm b}/t_{\rm E}$) while SS envelope should be negligible in the ``slow-piston" regime ($t_{\rm spread} \gg t_{\rm E}$). However, they also claim that in realistic simulations, $t_{\rm spread} \sim (0.5-1) t_{\rm E}$ is more reasonable to enable the wave-like structure to carry enough kinetic energy. In our fiducial simulations $t_{\rm spread} \sim 2 t_{\rm E} (t_{\rm E} \sim 7~{\rm Myr})$. Hence our fiducial feedback is well into a ``slow-piston" regime. We do not expect any strong shocks in our simulated ICM. We carry out several low/medium resolution simulations (e.g., $M_{\rm 0.5t_E, *}$, $L_{\rm 0.5t_E, *}$ in Table \ref{tab:sims}) in which $t_{\rm spread}/t_{\rm E} = 0.5$. For a detailed description of our parameter range see section \ref{sec:desc_tab}.

\subsection{Dissipation of energy in a viscous intracluster medium}
As energy is injected into the cluster core over a time $t_{\rm spread}$, the density lowers and the region ($<R_{\rm inj}$) expands. Thermal energy is converted to mechanical energy in the process and sound waves are triggered. The kinetic energy of the waves must be dissipated at larger radii. We can write the rate of change of kinetic energy for this second conversion in the following way using :
\begin{eqnarray}
\label{eq:ke1}
\nonumber
\frac{\partial \Big(\frac{1}{2} \rho u^2\Big)}{\partial t} &=& u_i \Big(\frac{1}{2}u_i \frac{\partial \rho }{\partial t} + \rho \frac{\partial  u_i}{\partial t} \Big) \\
\nonumber
&=& u_i \Big(-\frac{1}{2}u_i [{\partial}_j(\rho u_j)] - \rho [u_j {\partial}_j u_i \\
\nonumber
&+& \frac{1}{\rho}({\partial}_j {\delta}_{ij} p - {\partial}_j {\Pi}_{ij}) + {\partial}_j {\psi}_{\rm NFW}] \Big) \\
\nonumber
= u_i \Big({-\partial}_j(\rho u_j u_i) &+& \frac{1}{2}u_i [{\partial}_j(\rho u_j)] - {\partial}_i p + {\partial}_j {\Pi}_{ij} \\
&-& \rho {\partial}_j {\psi}_{\rm NFW}\Big)
\end{eqnarray} 
Now, ${\partial}_j[\rho u_j \frac{1}{2}u_i^2] + \frac{1}{2} u_i^2 {\partial}_j(\rho u_j) = u_i {\partial}_j (\rho u_i u_j)$ and collecting the flux terms we can write eq. \ref{eq:ke1} as
\begin{eqnarray}
\label{eq:ke2}
\nonumber
\frac{\partial \Big(\frac{1}{2} \rho u^2\Big)}{\partial t} &=& - {\partial}_i \Big(\rho u_i [\frac{1}{2}u^2 + \frac{p}{\rho} + \psi_{\rm NFW}] - u_j{\Pi}_{ij}\Big)\\
\nonumber
&-& {\Pi}_{ij}{\partial}_ju_i - p{\partial}_iu_i 
+ {\psi}_{\rm NFW} {\partial}_i(\rho u_i)
\end{eqnarray} 

The first term in the RHS of eq. \ref{eq:ke2} is the flux of energy including the viscous flux. The rest of the terms account for the gain/loss of kinetic energy aided by internal and external forces. The viscous dissipation rate is given by:
\begin{eqnarray}
\label{eq:vdiss}
{\varepsilon}_{\rm diss} = {\Pi}_{ij}{\partial}_ju_i
\end{eqnarray} 

\subsection{Characteristic timescales}
\label{sec:timescales}
There are four characteristic timescales of the problem: $t_{\rm cool}$, $t_{\rm visc}$, $t_{\rm visc, diss}$ and $t_{\rm BV} (1/N_{\rm BV}$; explained in section \ref{sec:th_sw}). The energy injection in each episode is distributed uniformly across $t_{\rm spread}$.
The episodes of gentle feedback occur at every $t_{\rm duty} = 100$ Myr (fiducial). We define cooling timescale as
\begin{eqnarray}
t_{\rm cool} = \frac{p}{(\gamma -1)n_e n_i \Lambda(T)} 
\end{eqnarray} 
We will sometime use `$0$' in the subscript of $t_{\rm cool}$ to denote values at $t=0$. Note that in a thermally balanced ICM, the growth timescale of isobaric thermal instability is,
\begin{eqnarray}
t_{\rm TI} = \frac{\gamma t_{\rm cool}}{2 - \Lambda_{\rm T}} 
\end{eqnarray} 
Here $\Lambda_{\rm T} = \partial \ln \Lambda/ \partial T$. Since the observed galaxy clusters appear to be very close to thermal equilibrium, this timescale is often evoked for cold gas formation in the cores. The diffuse hot gas is at $\gtrsim 10^7$ K at which isobaric thermal instability is expected.

We also have a viscous timescale given by,
\begin{eqnarray}
t_{\rm visc} = \frac{L^2}{\nu}
\end{eqnarray}
where $L$ is a length scale appropriate for the problem. Note that we do not expect a preferred length scale in the ICM since it is a complex, dynamic medium. However, in the subsequent section (section \ref{sec:th_sw}), we will deduce that given a length scale $L$, small-amplitude sound waves (also linear g-modes) may decay at the viscous timescale given above. Thus it is probably obvious that short wavelength sound waves decay fast within a short distance from the source. A characteristic viscous dissipation timescale is,
\begin{eqnarray}
\label{eq:vdiss_t}
t_{\rm visc, diss} = {\Big(\rho\frac{\nu}{{\varepsilon}_{\rm diss}}\Big)}^{\frac{1}{2}}
\end{eqnarray}
where ${\varepsilon}_{\rm diss}$ is the rate of viscous dissipation (eq \ref{eq:vdiss}). Note that this timescale is not explicitly dependent on the temperature of the gas. Non-linear flows (e.g., small eddies) in the medium are expected to viscously dissipate at this timescale. Since sound waves maintain small amplitudes (linearity) relative to isobaric perturbations until there is significant steepening in the outskirts of the core, we expect incompressible fluctuations to dissipate at $t_{\rm visc, diss}$.

\subsection{Diagnostics}
\label{sec:diag}
\subsubsection{Flux and spatial averaging}
In this section we describe the most important methods of averaging and analysing the data obtained from simulations. Firstly, we note that any elemental area of magnitude $r^2 \sin \theta d\theta d\phi$ is along $\hat{r}$ and in order to calculate area averages or outgoing fluxes we need an area vector in the cartesian geometry. 
\begin{eqnarray}
\nonumber
d\mathbf{S_{\rm cart}} = r^2 sin \theta [\hat{d \theta} \times \hat{d \phi}] = \frac{x dy dz}{r} \hat{x}  + \frac{y dz dx}{r} \hat{y} + \frac{z dx dy}{r} \hat{z} 
\end{eqnarray}
where $x,y,z, r$ are cluster-centered variables in the computational box and the magnitude of the area is simply $dS_{\rm cart} = dx^2$ if $dx=dy=dz$. In order to calculate the radial flux, we use the elemental area vector. For a scalar variable $Q (x,y,z)$ and a vector $\mathbf{V}(x,y,z)$, we compute the surface average and flux in the following way:
\begin{eqnarray}
\langle Q \rangle (r) = \frac{\int Q dS_{\rm cart}}{\int dS_{\rm cart}} \\
F = \int \mathbf{V} \cdot d \mathbf{S_{\rm cart}}
\end{eqnarray}
For a range of $r$ between $0$ and $x_{\rm max}$ (note we get values of $r$ only upto $\sqrt{3}x_{\rm max}/2$ since $x_{\rm max}$ is the diameter of the largest sphere in the cubic box), we define $N_{\rm r}$ shells and integrate across each of the radial shells defined by that grid. $N_{\rm r}$ is constant across most simulations. Note that for the highest resolution $H_{\rm f}$, we also tested with double the number (thinner) of shells since ideally for flux calculation the shell width should be infinitesimal.\footnote{If the shells are too wide, a higher resolution simulation may acquire too many grid cells (in radial direction) and thus the flux can be somewhat higher for a shell with same width compared to a lower resolution. This effect will be insignificant if shells are thin enough. However, thinner shells will obviously generate a smaller flux than physically wider shells. The ideal scenario for flux calculation is extremely high resolution and extremely thin shells.}

The fluctuations in a variable $Q$ is
\begin{eqnarray}
\delta Q (x,y,z) = Q(x,y,z) - \langle Q \rangle (r)
\end{eqnarray}
Note that the number of radial points or shells considered for this analysis, may introduce small-amplitude static (not physically evolving, but proportional to the background) spatial features in the fluctuations. These features occur since we map a cartesian geometry to a spherical geometry. Such high-frequency features are filtered out for any spatial fluctuation maps shown/used in this work. 

\subsubsection{Helmholtz decomposition}
\label{sec:decomp}
The weak-shock injected into the ICM is expected to trigger both compressible and incompressible velocity modes. Hence in order to distinguish sound waves we perform Helmholtz decomposition on the velocity field (similar to section 4.3 in \citealt{2015ApJ_reynolds}) in the following way:
\begin{eqnarray}
\nonumber
\mathbf{u} = \mathbf{u}_{\rm c} + \mathbf{u}_{\rm ic} \\
\nonumber
\nabla \cdot \mathbf{u} = \nabla \cdot \mathbf{u}_{\mathrm{c}} + \nabla \cdot \mathbf{u}_{\mathrm{ic}}\\
\nonumber
\nabla \cdot \mathbf{u} = {\nabla}^2 {\phi}_{\rm c} \\
\nonumber
{\phi}_{\rm c, FT} = -i\mathbf{k} \cdot \mathbf{u}_{\rm FT}/ k^2
\end{eqnarray}
where FT refers to `Fourier transform', $\mathbf{u}_{\rm c}$ and $\mathbf{u}_{\rm ic}$ refer to the compressible and incompressible components, and the curl-free component $\mathbf{u}_{\rm c} = \nabla \phi_{\rm c}$. We distinguish between compressible sound waves and solenoidal motions using this method. 

\subsection{Theoretical preliminaries: local linear sound waves}
\label{sec:th_sw}
Here we will discuss how sound waves and g-modes behave locally in a viscous medium maintained at thermal equilibrium. The interactions of these waves with the ambient medium is crucial to understand how a gentle feedback may or may not sustain a long-term prevention mechanism against cooling-flows. Primarily, we motivate an intuitive understanding from the simplest physical scenario in which sound waves propagate in an uniform viscous medium.

\subsubsection{No stratification}
\label{sec:sw_simplest}
Let us assume sound waves are propagating in an uniform 1D medium of density, pressure, temperature, etc denoted by `0' suffix, ${\rho}_0$, $p_0$, and $T_0$ respectively. For any of these variables, F, we assume that the fluctuations are small ($\delta F/F_0 \ll 1$ where $F = F_0 + \delta F$). In the absence of stratification, there are only sound waves generated in the medium. We take a WKB approximation in 1D such that the fluctuations (Eulerian) take a form $\delta F = F^{\prime} e^{i(kx - \omega t)}$ where $k$ denotes the wave-number and $\omega$ denotes the oscillation rate of the fluctuation. The linearised hydrodynamic equations (ignoring second-order terms) in terms of the $k$ and $\omega$ are,
\begin{eqnarray}
-i \omega {\rho}^{\prime} &=& - ik{\rho}_0 u^{\prime}\\
-i \omega {\rho}_0 u^{\prime} &=& -ikp^{\prime} + ({ik})^2 {\eta}_{\rm c} T^{\frac{5}{2}}_{70} u^{\prime}
\end{eqnarray} 
where $\eta_{\rm c} = \frac{4}{3} \xi_{\nu} \mu m_{\rm p} {10}^{25} = 1.38 (\xi_{\nu} = 0.1)$ is a constant associated with the viscosity. In eq \ref{eq:eq3}, the last term adds second-order terms in perturbation. Hence the energy equation is simply, $p^{\prime}/p_0 = \gamma {\rho}^{\prime}/{\rho}_0 = c^2_{\rm s0} {\rho}^{\prime}$. Replacing ${\rho}^{\prime}$ from first eq in the second and subsequently multiplying it by $i\omega/{\rho}_0$,
\begin{eqnarray}
\nonumber
-i \omega {\rho}_0 u^{\prime}&=& -ikc^2_{\rm s0} \Big( \frac{k {\rho}_0 u^{\prime}}{\omega}\Big) + ({ik})^2 {\eta}_{\rm c} T^{\frac{5}{2}}_{70} u^{\prime}\\
\label{eq:dispnostr}
\implies {\omega}^2 &=& k^2 c^2_{\rm s0} - i \omega {k}^2 \frac{{\eta}_{\rm c} T^{\frac{5}{2}}_{70} }{{\rho}_0} \\
\label{eq:knostr}
\implies k &=& \pm \frac{\omega/c_{\rm s0}}{\sqrt{1 - i \delta_{\nu}}}
\end{eqnarray} 
where $\delta_{\nu} = \frac{\omega {\eta}_{\rm c}T^{\frac{5}{2}}_{70} }{{\rho}_0 c^2_{\rm s0}}$ is dimensionless (similar to the last term in eq 207 in \citealt{lighthill2001waves}). Hence $\omega_{\nu} = \frac{{\rho}_0 c^2_{\rm s0}}{{\eta}_{\rm c}T^{\frac{5}{2}}_{70} }$ is a characteristic frequency scale. Note that another way of rewriting the dispersion relation is,
\begin{eqnarray}
\label{eq:omegnostr}
\omega = -\frac{i}{2} k^2 {\nu}_{\rm c} \pm kc_{\rm s0} \sqrt{1 - k^2 l^2_{\nu}}
\end{eqnarray}
which again provides interesting scales, ${\nu}_{\rm c} = \frac{{\eta}_{\rm c}T^{\frac{5}{2}}_{70} }{{\rho}_0 }$ and $l_{\nu} =\frac{{\eta}_{\rm c}T^{\frac{5}{2}}_{70}}{2{\rho}_0 c_{\rm s0}}$ where $c_{\rm s0} l_{\nu} = {\nu}_{\rm c}/2$. This above equation comes if we solve eq \ref{eq:dispnostr} as a quadratic equation in $\omega$. 

The significant aspects of eqs \ref{eq:knostr} and \ref{eq:omegnostr} are about how sound waves get affected. The decay rate with time is the real part of $-i \omega$ ($-k^2 \nu_{\rm c}/2$). At appropriate regimes, $k \ll 1/l_{\nu}$, we arrive at the wave properties of sound waves unaffected by viscosity ($c_{\rm s0} = \omega/k$). All the scales are relevant to understand how this contributes to the gas evolution, e.g $l_{\nu}$ for hot dense gas at ${\rho}_0 = {10}^{-25}~{\rm g{cm}^{-3}}$, $c_{\rm s0} = 471~{\rm km/s}$, and $T_{70} = 1(10^7~{\rm K})$, is $\sim 0.05~{\rm pc}$.  

Note that for $k= 1/l_{\nu}$ or $k \gg 1/l_{\nu}$, there is no oscillation/propagation. The latter condition further means, 
\begin{eqnarray}
\nonumber
-i\omega &=& -\frac{1}{2} k^2 {\nu}_{\rm c} \mp i^2k^2c_{\rm s0} \sqrt{l^2_{\nu} - 1/k^2}\\
\nonumber
&=& 0 / -k^2 \nu_{\rm c}
\end{eqnarray}
Below $l_{\nu}$ scale, dissipation is inevitable. Above this scale, wave speed can be reduced significantly until $1/k \gg l_{\nu}$ which is the limit of linearly propagating sound waves in a medium unaffected by viscosity.

Note that the sound crossing time of twice this characteristic length ($2l_{\nu}$) is $t_{\nu}=1/{\omega}_{\nu}$ ($c_{\rm s0}t_{\nu} = 2 l_{\nu}$) where according to our earlier definition ${\delta}_{\nu} = \omega/{\omega}_{\nu}$. A general form for a characteristic attenuation length scale (using eqs \ref{eq:knostr} as before) is $l_{\rm a} = 2c_{\rm s0}/\omega {\delta}_{\nu}$ (see first term in eq 79.6 in \citealt{LANDAU1987251}). For any fluctuation with frequencies $\omega \sim {\omega}_{\nu}$, this implies, $l_{\rm a} \sim 2c_{\rm s0}/{\omega}_{\nu} = 4 l_{\nu}$. With $\omega \gg {\omega}_{\nu}$ ($\omega \ll {\omega}_{\nu}$), $l_{\rm a}$ is much smaller (larger) than $4l_{\nu}$. To reiterate, high (low) frequency waves get attenuated within a short (long) distance from source.

\subsubsection{Cooling ($l_{\rm cool}$) and viscous ($l_{\nu}$) length scales for sound waves}
\label{sec:sw_cool}
Since sound waves are almost isentropic, these are not expected to be thermally unstable. But sound waves can be slightly heated in the compression phase which may lead to instability (eq 5 in \citealt{1965ApJ_field}). If we consider the high-frequency solution of the acoustic and gravity modes in presence of thermal instability (section 3.2 \citealt{2016MNRAS_choudhury} with perturbations $\propto e^{\sigma t}$), we get,
\begin{eqnarray}
{\sigma}^2 + \sigma \frac{d \ln \Lambda/d \ln T}{t_{\rm cool}} + c^2_{\rm s0} k^2_{\rm R}\\
\nonumber 
\sigma = - \frac{\Lambda_{\rm T}}{2t_{\rm cool}} \pm i c_{\rm s0} k_{\rm R} \sqrt{1 - \frac{{\Lambda}^2_{\rm T}}{4 t^2_{\rm cool}c^2_{\rm s0} k^2_{\rm R}}}
\end{eqnarray}
where ${\Lambda}_{\rm T} = d \ln \Lambda/d \ln T$ and putting $l_{\rm cool} = \frac{2t_{\rm cool}c_{\rm s0}}{\vert \Lambda_{\rm T}\vert}$ we get a length scale such that for $k_{\rm R}>> 1/l_{\rm cool}$ sound waves are almost unaffected by cooling. Above this scale, $k_{\rm R} << 1/l_{\rm cool}$ sound waves cannot propagate and decay for positive $\Lambda_{\rm T}$ (Brehmmstrahlung regime) and may grow for negative $\Lambda_{\rm T}$ (at $\sim 10^{5-6}$ K). In the previous section, for hot dense gas at ${\rho}_0 = {10}^{-25}~{\rm g{cm}^{-3}}$, $c_{\rm s0} = 471~{\rm km/s}$, and $T_{70} = 1(10^7~{\rm K})$, we show that is $l_{\nu} = 0.05~{\rm pc}$. Additionally, using $p_{\rm thermal} \sim 10^{-10} {\rm g{cm}^2 s^{-2}}$, $\Lambda_{\rm T} = 0.5$ ($\Lambda = \Lambda_0 T^{0.5}$), $l_{\rm cool} = 4t_{\rm cool} c_{\rm s0} \sim 400$ kpc. In between these two scales ($0.05~{\rm pc} < l_{\rm scale} < 400~{\rm kpc}$), linear sound waves are expected to propagate unhindered. In reality, waves may steepen in the outskirts of clusters.

\subsubsection {With stratification}
We include the details of the linear equations in Appendix \ref{sec:applin}. The viscous ICM is a favourable host for dissipating the waves. We will use eq \ref{eq:fulldisp} to understand the limiting cases of viscous effects on the modes mentioned above. Firstly, note that for a low frequency oscillation with small viscous effects (ignoring high powers of $\bar{\nu}$), we get $-i \omega = - \bar{\nu} B_{2\nu} = -\bar{\nu} (\frac{4}{3}l^2 + kl)$, or a purely decaying mode. Secondly, let us focus on high-frequency sound waves: 
\begin{eqnarray}
\nonumber
{(i \omega)}^2 - {(i \omega)} (2 \bar{\nu}B_{2\nu} +\bar{\nu}A_{1\nu}) &+&  \Big[{\bar{\nu}}^2(B^2_{2\nu}\\
\nonumber
+ 2 A_{1\nu}B_{2\nu} - A_{2\nu}B_{1\nu}) &+& c^2_{\rm s0}k^2_{\rm R} - ikg \Big(1+ \frac{c^2_{\rm s0}N^2_{\rm BV}}{g^2} \Big)\Big]\\
 &=& 0
\end{eqnarray}
If we ignore the acoustic cut-off/attenuation in the above (imaginary terms in the coefficients of the quadratic equation), it provides a pair of decaying, modified sound waves: 
\begin{eqnarray}
\nonumber
-i{\omega} &=& -\bar{\nu}(B_{2\nu}+ A_{1\nu}/2)\\
&\pm& ic_{\rm s0}k_{\rm R}\sqrt{1- \frac{{\bar{\nu}}^2 F(A_{1\nu}, A_{2\nu}, B_{1\nu}, B_{2\nu})}{c^2_{\rm s0}k^2_{\rm R}}}
\end{eqnarray}
where $F(A_{1\nu}, A_{2\nu}, B_{1\nu}, B_{2\nu})$ is a function of $l$ and $k$ such that $F \propto 1/l^4_{\rm scale}$ where $l_{\rm scale}$ is a length by dimension, formed by combining $l$ and $k$. Thus, for a moderately large scale sound wave ($l^4_{\rm scale} k^2_{\rm R} \gg \frac{{\bar{\nu}}^2}{c^2_{\rm s0}}$), the propagation is unhindered (similar to the uniform viscous medium in section \ref{sec:sw_simplest}). 

\subsection{Summary of the linear sound waves}
\label{sec:summ_sw}
Here we will summarize the important characteristics that small-amplitude sound waves in viscous stratified ICM may have, based on the detailed linear theory above:
\begin{itemize}
    \item {\bf Linear sound waves of long wavelengths are least affected by viscosity:} High-frequency ($\omega \gg N_{\rm BV}$) sound waves are not expected to interact with g-modes out to large distances. These can be attenuated either beyond a pressure scale height (see Appendix \ref{sec:applin} and Figure \ref{fig:sflux}) or by viscosity. If the length-scales of fluctuations are $\sim \bar{\nu}/c_{\rm s0}$, the propagation of sound is delayed by viscosity ($\omega <kc_{\rm s}$). This length scale is dynamically varying for different physical conditions but typically in the sub-parsec to a kpc range. Note that the decay rate of the wave at such length scale is $\sim \bar{\nu}/l^2_{\rm scale}$, or in other words, small-scale waves will decay the fastest. Long wavelengths have a greater chance of survival against decay and attenuation.
    \item {\bf Cooling may result in decay of sound waves of very long wavelengths:} Sound waves, in a globally thermally stable core, decay if cooling time is significantly short; in addition the propagation is delayed as well (section \ref{sec:sw_cool}). In cluster cores, the shortest cooling times occur in the innermost regions ($\sim$ kpc). Hence a fraction of the sound waves generated by the central engine may radiate and decay without escaping the core after a few $t_{\rm cool}$. Note that for our simulation the smallest $t_{\rm cool}$ is $\sim 700$ Myr and hence this effect may not be significant. 
    \item {\bf Gravity waves may brake sound waves if timescales are comparable:}The moderate frequency sound waves ($k_{\rm R} c_{\rm s0} \sim N_{\rm BV}$ and $k_{\rm R} c_{\rm s0} \gtrsim N_{\rm BV}$) are slowed down in presence of g-modes. Large-scale modes along gravity ($k<<l$) may grow/decay as well (see Appendix \ref{sec:applin}). This is a rare occurence for the ICM core in which sound timescale is much shorter than gravity wave timescale.
\end{itemize}

\section{Results}
\label{sec: res}
\begin{figure*}
    \includegraphics[width=18cm]{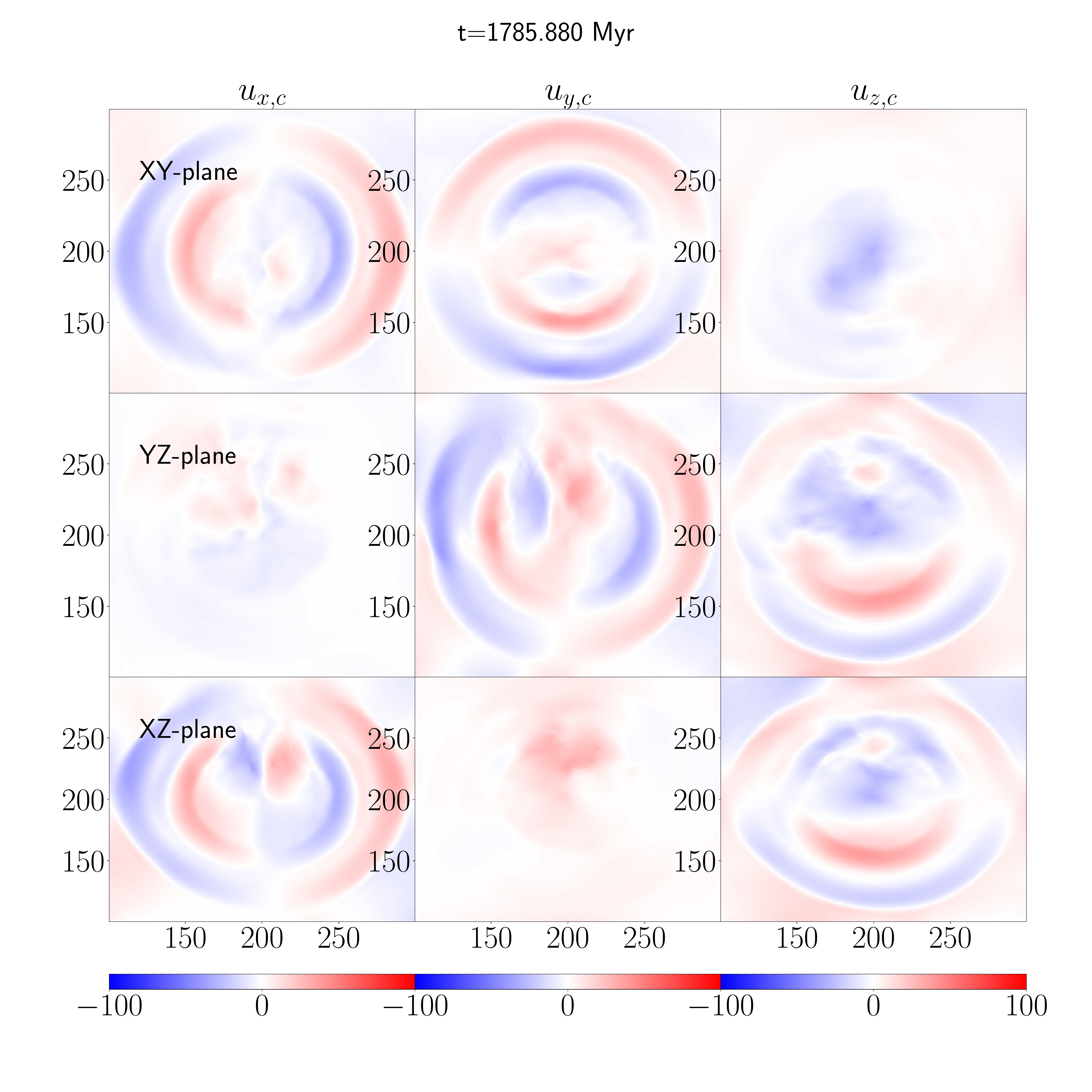}
    \centering
    \caption{The compressible velocity map in the mid-planes of the inner uniform cube on which the Helmholtz decomposition is done for $M_{\rm f, HE}$. The large scale spherical sound wave propagates out and the inner regions are weakly shocked. (In $M_{\rm 0.5t_E, HE}$ we find similar weakly shocked central region and much stronger outgoing waves. This is a \href{https://youtu.be/K_Nzw3IDwj8}{link} to the animation.).}
    \label{fig:compre_v}
\end{figure*}

\begin{figure*}
    \includegraphics[width=18cm]{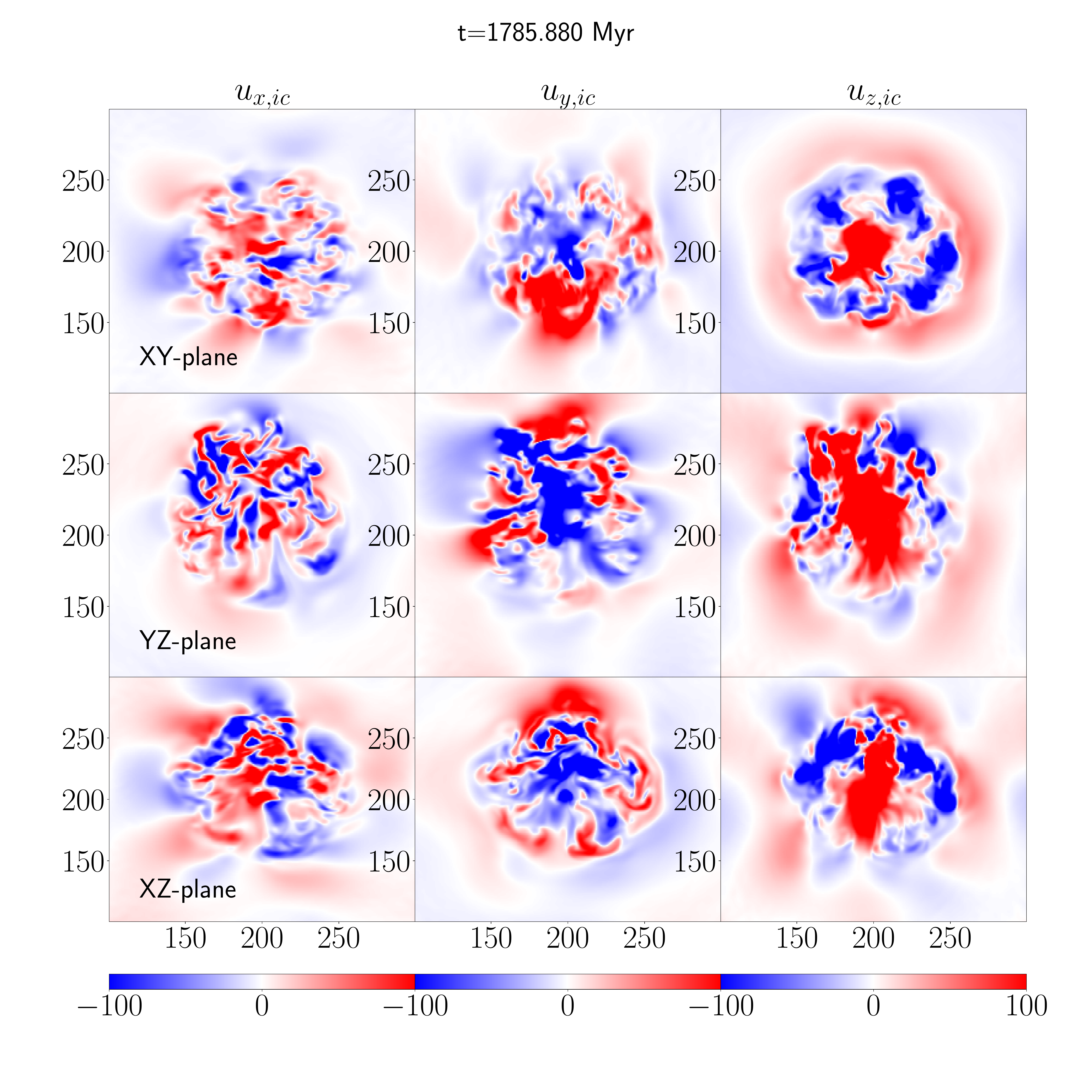}
    \centering
    \caption{The incompressible velocity map in the mid-planes of the inner uniform cube on which the Helmholtz decomposition is done for $M_{\rm f, HE}$. The central $\rm 50$ kpc is dominated by incompressible dynamics. The vorticity map mirrors the swirling motions shown here. The large scale alternate velocity features are possibly large-scale g-modes.}
    \label{fig:incompre_v}
\end{figure*}

\begin{figure*}
    \includegraphics[width=18cm]{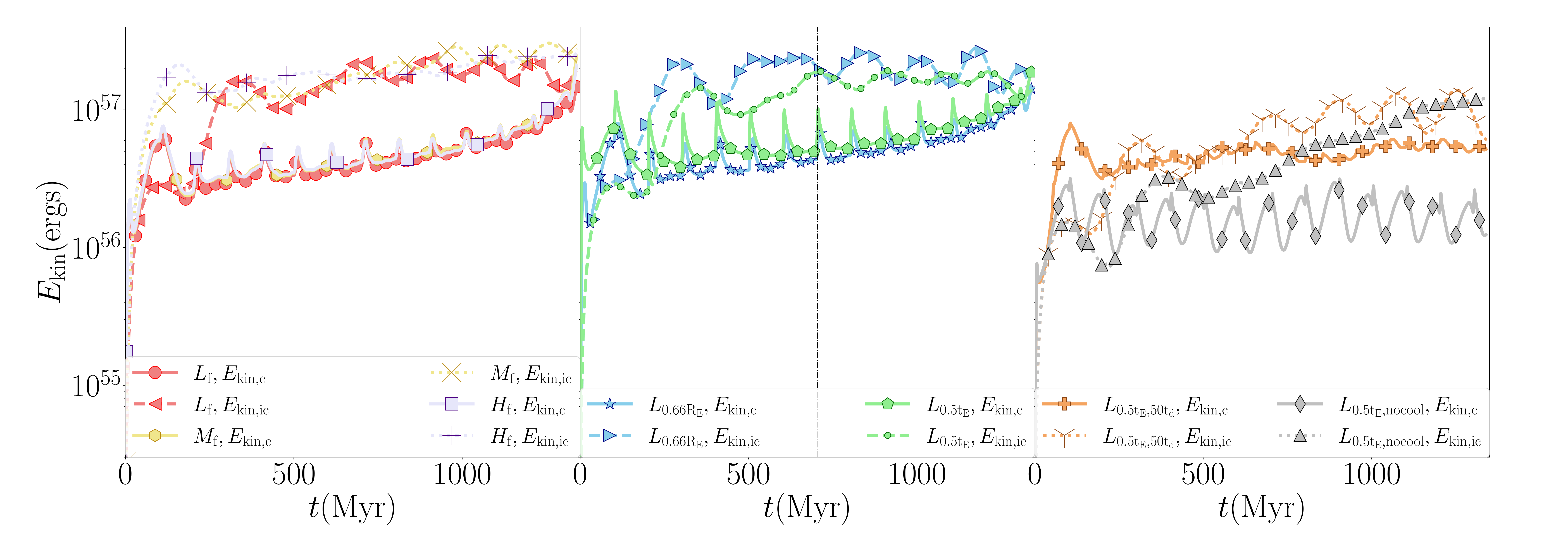}
    \centering
    \caption{The kinetic energy integrated over the inner uniform cube of the simulation box. The solid lines represent compressible mode $E_{\rm kin, c} = \int \rho u^2_{\rm c} dV$ and dashed lines represent incompressible mode $E_{\rm kin, ic} = \int \rho u^2_{\rm ic} dV$. The legend uses the simulation names from Table \ref{tab:sims}. The vertical dashed line denotes a chosen time when the kinetic energies in $L$-simulations are close (not equal). Also, check Figure \ref{fig:KE_extend} for a brief discussion on how viscosity dynamically affects this result.}
 \label{fig:KE}
\end{figure*}   

\begin{figure*}
    \includegraphics[width=15cm]{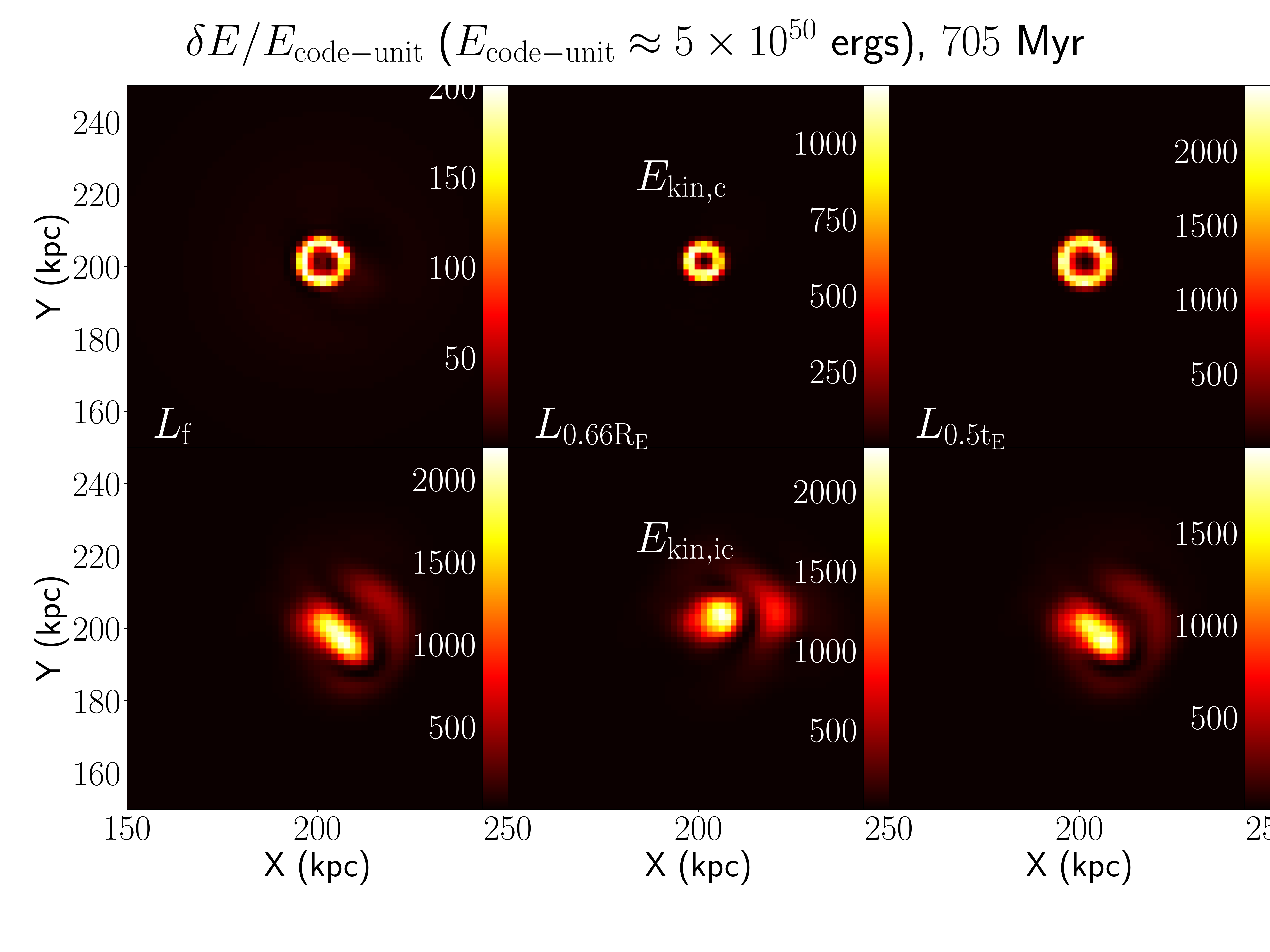}
    \centering
    \caption{The kinetic energy components $\delta E_{\rm kin, c} = \frac{1}{2}\rho u^2_{\rm c} dV$ and $\delta E_{\rm kin, ic} = \frac{1}{2}\rho u^2_{\rm ic} dV$ for $L_{\rm f}$, $L_{\rm 0.66R_E}$ and $L_{\rm 0.5t_E}$. The exact time corresponding to this slice ($705$ Myr) is not uniquely interesting, but chosen due to less variation in the total kinetic energies in Figure \ref{fig:KE}. Note that the time corresponds to a peak in total $E_{\rm kin, c}$ for $L_{\rm 0.5t_E}$.}
 \label{fig:KE_spatial}
\end{figure*}   
\subsection{Description of the parameter space in the suite of simulations in Table \ref{tab:sims}}
\label{sec:desc_tab}
In this section we motivate the range of various parameters in Table \ref{tab:sims} which includes our suite of simulations. In section \ref{sec:ws_lit}, we mention the net energy injection in two observed clusters to be $E_{\rm inj, M87} = 5.03 \times 10^{57}$ ergs and $E_{\rm inj, Perseus} = 9.67 \times 10^{58}$ ergs. We use two values of $E_{\rm inj} = [10^{58} ({\rm f}), 10^{59}({\rm f, HE})]$ ergs from this observed range in our simulation suite. The first case is used to test convergence ([$L_{\rm f}, M_{\rm f}, H_{\rm f}$] are simulations in the increasing order of resolutions). We see reasonable convergence in spatial features between $M_{\rm f}$ and $H_{\rm f}$ hence `${\rm f, HE}$' is carried out at medium ($M_{\rm f, HE}$) resolution only. The fiducial (`${\rm f}$') $t_{\rm spread}/t_{\rm E} = 2$ is in the slow-piston or gentle feedback regime. We explore a moderately strong feedback with $t_{\rm spread}/t_{\rm E} = 0.5$ (denoted by subscript `$0.5 t_{\rm E}$' instead of `${\rm f}$'). We do not explore extremely strong feedback since that is not our objective. Our initial condition is modeled based on the radial temperature profile of Perseus. Hence the first fiducial case ($E_{\rm inj} \sim 2E_{\rm inj, M87} < E_{\rm inj, Perseus}$) possibly fails to prevent cooling flow after $\sim (1-1.5) t_{\rm cool, min}$ while the second fiducial case (`${\rm f,HE}$') and the second moderately strong feedback case (`${\rm 0.5 t_E,HE}$') prevents it longer successfully. All other cases have appropriate subscripts and parameters as noted in Table \ref{tab:sims}.

\begin{table*}
    \caption{Table of simulations}
	\label{tab:sims}
    \begin{tabular}{cccccccccccc}
    \hline
    \multicolumn{1}{|p{1cm}|}{\centering Simulation\\name} & 
    \multicolumn{1}{|p{1.5cm}|}{\centering Full \\resolution (Inner-cube)} & 
    \multicolumn{1}{|p{1.5cm}|}{\centering ${dt}_{\rm dump}$\\(data dumping)} &
    \multicolumn{1}{|p{1cm}|}{\centering $E_{\rm inj}$\\(energy injected in ergs)} &
    \multicolumn{1}{|p{1cm}|}{\centering $t_{\rm spread}$\\(duration in Myr)} & 
    \multicolumn{1}{|p{1cm}|}{\centering $P_{\rm inj} = E_{\rm inj}/t_{\rm spread}$\\(erg/s)} &
    \multicolumn{1}{|p{1cm}|}{\centering $R_{\rm inj}$\\(in kpc)} & 
    \multicolumn{1}{|p{1cm}|}{\centering $t_{\rm duty}$\\ (repetition time in Myr)} &
    \multicolumn{1}{|p{1cm}|}{\centering $R_{\rm inj}/R_{\rm E}$} &
    \multicolumn{1}{|p{1cm}|}{\centering $t_{\rm spread}/t_{\rm E}$}&
    \multicolumn{1}{|p{1cm}|}{ other comments}\\
    \hline
     $L_{\rm f}$&${192}^3$(${128}^3$)&$0.001(\sim 1$Myr)&${10}^{58}$&$14$&$2.3\times{10}^{43}$&$5.8$&$100$&$0.92$&$2$& \\
     $M_{\rm f}$&${384}^3$(${256}^3$)&$0.003(\sim 3$Myr)&${10}^{58}$&$14$&$2.3\times{10}^{43}$&$5.8$&$100$&$0.92$&$2$& \\
     $M_{\rm f, nv}$&${384}^3$(${256}^3$)&$0.003(\sim 3$Myr)&${10}^{58}$&$14$&$2.3\times{10}^{43}$&$5.8$&$100$&$0.92$&$2$& no viscosity \\
     $H_{\rm f}$&${768}^3$(${512}^3$)&$0.003$&${10}^{58}$&$14$&$2.3\times{10}^{43}$&$5.8$&$100$&$0.92$&$2$& \\
     $L_{\rm 0.66R_E}$&${192}^3$(${128}^3$)&$0.001$&${10}^{58}$&$14$&$2.3\times{10}^{43}$&$4.1$&$100$&$0.66$&$2$& \\
     $L_{\rm 0.5t_E}$&${192}^3$(${128}^3$)&$0.001$&${10}^{58}$&$3.5$&$9.2\times{10}^{43}$&$5.8$&$100$&$0.92$&$0.5$& \\
     $L_{\rm 0.5t_E, nv}$&${192}^3$(${128}^3$)&$0.001$&${10}^{58}$&$3.5$&$9.2\times{10}^{43}$&$5.8$&$100$&$0.92$&$0.5$&no viscosity \\
     $L_{\rm 0.5t_E, HLLC}$&${192}^3$(${128}^3$)&$0.001$&${10}^{58}$&$3.5$&$9.2\times{10}^{43}$&$5.8$&$100$&$0.92$&$0.5$&HLLC \\
     $L_{\rm cf}$&${192}^3$(${128}^3$)&$0.001$&$-$&$-$&$-$&$-$&$-$&$-$&$-$&cooling-flow \\
     $L_{\rm 0.5t_E,nocool}$&${192}^3$(${128}^3$)&$0.001$&${10}^{58}$&$3.5$&$9.2\times{10}^{43}$&$5.8$&$100$&$0.92$&$0.5$& \\
     $L_{\rm 0.5t_E, 50t_d}$&${192}^3$(${128}^3$)&$0.001$&${10}^{58}$&$3.5$&$9.2\times{10}^{43}$&$5.8$&$50$&$0.92$&$0.5$& \\
     $M_{\rm f, HE}$&${384}^3$(${256}^3$)&$0.003$&${10}^{59}$&$30.3$&$1.1\times{10}^{44}$&$12.43$&$100$&$0.92$&$2$&HLLC \\
     $M_{\rm 0.5t_E, HE}$&${384}^3$(${256}^3$)&$0.003$&${10}^{59}$&$7.58$&$4.24\times{10}^{44}$&$12.43$&$100$&$0.92$&$0.5$&HLLC \\
    &&&&&&&& \\
    \hline
    \end{tabular}
\end{table*}

\subsection{Compressible and solenoidal velocities in the ICM}
\label{sec:v_comp_incomp}
We first use the inner uniform cube ($~100$ kpc radius) and decompose the three-dimensional velocity field into compressible and incompressible modes. Figures \ref{fig:compre_v} and \ref{fig:incompre_v} show the $u_{\rm c}$ and $u_{\rm ic}$ (as calculated by Helmholtz decomposition; see section \ref{sec:decomp}) for $M_{\rm f, HE}$. The former shows a quasi-spherical sound wave that has propagated across a radius of $\sim 100$ kpc while the inner regions are weakly shocked. The latter shows a turbulent inner core. In the outskirts of Figure \ref{fig:incompre_v}, we also see large-scale g-modes which break in the inner core. 

\subsubsection{How net $E_{\rm kin, c}$ and $E_{\rm kin, ic}$ evolve in various physical scenarios of low $E_{\rm inj}$ cases?}
Figure \ref{fig:KE} shows the compressible ($E_{\rm kin, c}$) and incompressible ($E_{\rm kin, ic}$) kinetic energy integrated over the inner data cube for all the low $E_{\rm inj}$ simulations in Table \ref{tab:sims}. It shows incompressible kinetic energy is always higher than the compressible kinetic energy (all dashed/dotted lines are above the solid lines for any given color) roughly by a factor of $2$. There are exceptions which we will discuss later. The peaks of the two kinetic energy components roughly correspond to the duty cycle of energy injection. We also find a moderate difference between $L_{\rm f}$ and $M_{\rm f}$ (red and yellow lines) and a minor difference between $M_{\rm f}$ and $H_{\rm f}$ (yellow and purple lines). This confirms that our medium resolution ($M_{*}$) simulations are very close to convergence.  

We also note two other interesting scenarios in Figure \ref{fig:KE}, $L_{\rm 0.66R_E}$ and $L_{\rm 0.5t_E}$. In the former, when the ejecta radius ($R_{\rm inj}$) is smaller than the fiducial $0.92R_{\rm E}$, the instantaneous power is $\sim 2.7$ times higher and both kinetic energy components are slightly enhanced at all times (compare red and blue lines). In the latter, when instantaneous power is larger by a factor of $4$, the incompressible energy remains almost unchanged and the compressible component shows sharp spikes at all the injection times (compare red and light green lines). This increase in the amount of kinetic energy carried by wave-like features (light green solid line versus red solid line) agrees with the expectations discussed in section \ref{sec:ws_lit}. This difference between cases $t_{\rm spread}/t_{\rm E}=2$ and $0.5$ is crucial to determine which energy carrier dominantly prevents cooling-flows. 

\citealt{2017MNRAS_tang} (eq 6) argues that the enthalpy of the ejecta steadily increases with decreasing energy injection rate. Consequently the escaping fraction carried by the wave reduces significantly. On the other hand, for higher rates of energy injection we progressively approach the strong shock regime in which thermalization predominantly occurs in the ejecta region due to shock heating. The transition phase from low to high rate of injection (e. g., $t_{\rm spread}/t_{\rm E} = 0.5$) can maximise the outgoing energy via waves. We will discuss further details on this using our medium resolution simulations ($M_{\rm f, HE}$ and $M_{\rm 0.5t_E, HE}$) in the following sections. 

In Figure \ref{fig:KE}, we also choose a vertical line denoting $705$ Myr, which passes through a peak in $E_{\rm kin,c}$ for $L_{\rm f}$, $L_{\rm 0.66R_E}$ and $L_{\rm 0.5t_E}$ (red, blue, green lines). At this time, we plot the XY spatial slice in each simulation (Figure \ref{fig:KE_spatial}). The spatial distribution of $E_{\rm kin, ic}$ is remarkably similar in all three cases. Regions of high $E_{\rm kin,c}$ in $L_{\rm 0.66R_E}$ are centrally confined compared to the other two cases as expected. It is interesting to note that $E_{\rm kin, c}$ is extremely high in $L_{\rm 0.5t_E}$. Evidently there is a better chance of generating more sound waves in `$0.5t_{\rm E}$' compared to the fiducial gentler feedback.

\begin{figure*}
    \includegraphics[width=18cm]{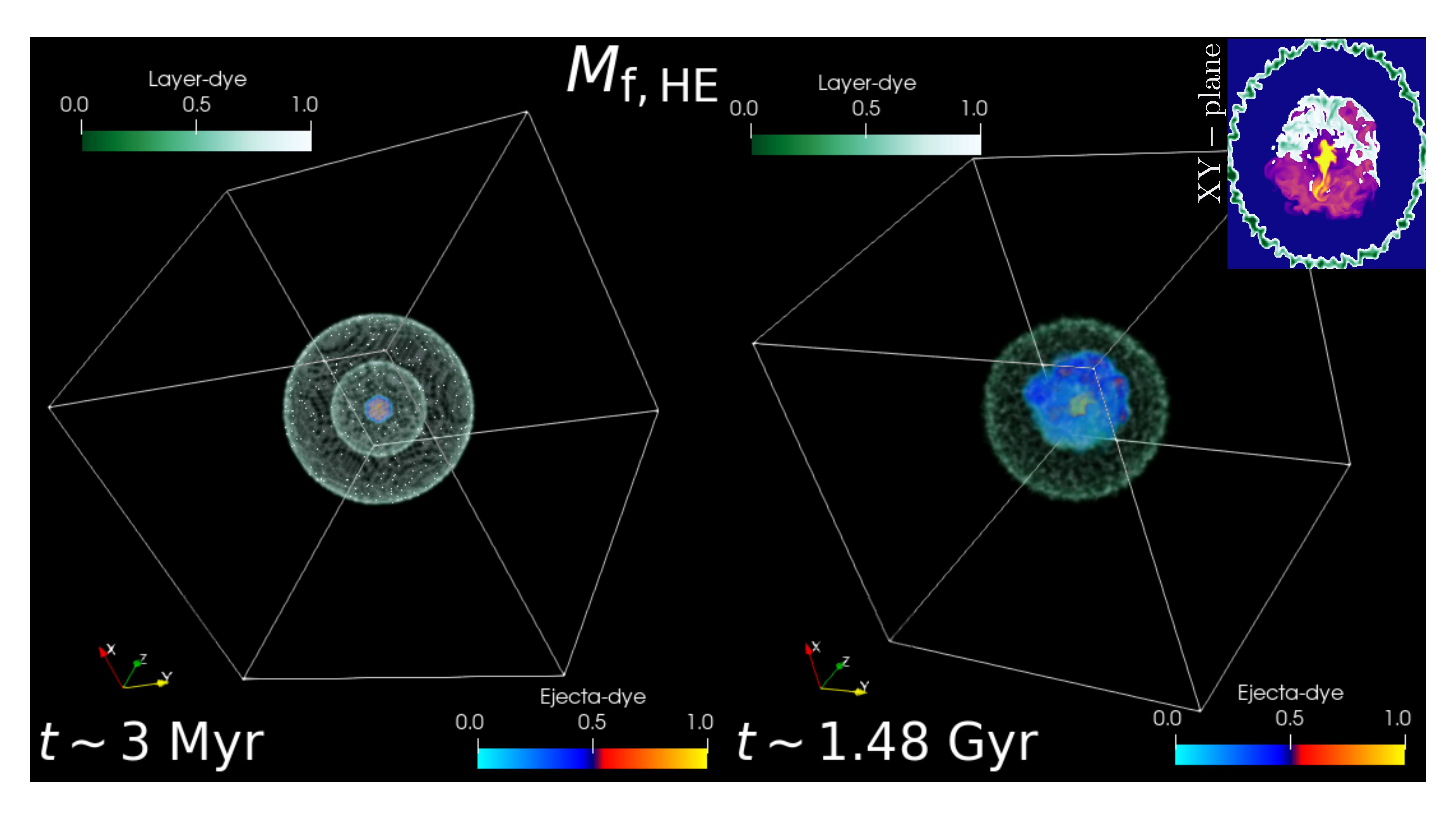}
    \centering
    \caption{The 3D volume rendering of two types of tracers injected in $M_{\rm f, HE}$ at different times. Left one is at $t\sim 3$ Myr and `Layer-dye' refers to the tracer injected at $t=0$ around $50(\pm 2)$kpc and $100(\pm 2)$kpc. Right one is at $t\sim 1.48$ Gyr and `Ejecta-dye' is injected along with energy within $R_{\rm inj}$ at every $t_{\rm duty}$. Note that the outer($100$ kpc) `Layer-dye' is well-preserved while the inner($50$ kpc) layer gets mixed with the `Ejecta-dye'. In the upper right corner we attach a zoom-in view of the XY-plane at $t\sim 1.48$ Gyr which shows the corrugations of the outer `Layer-dye' due to g-modes. In addition, it clearly shows how the inner layer is disrupted.}
 \label{fig:dye}
\end{figure*}   

\begin{figure*}
    \includegraphics[width=18cm]{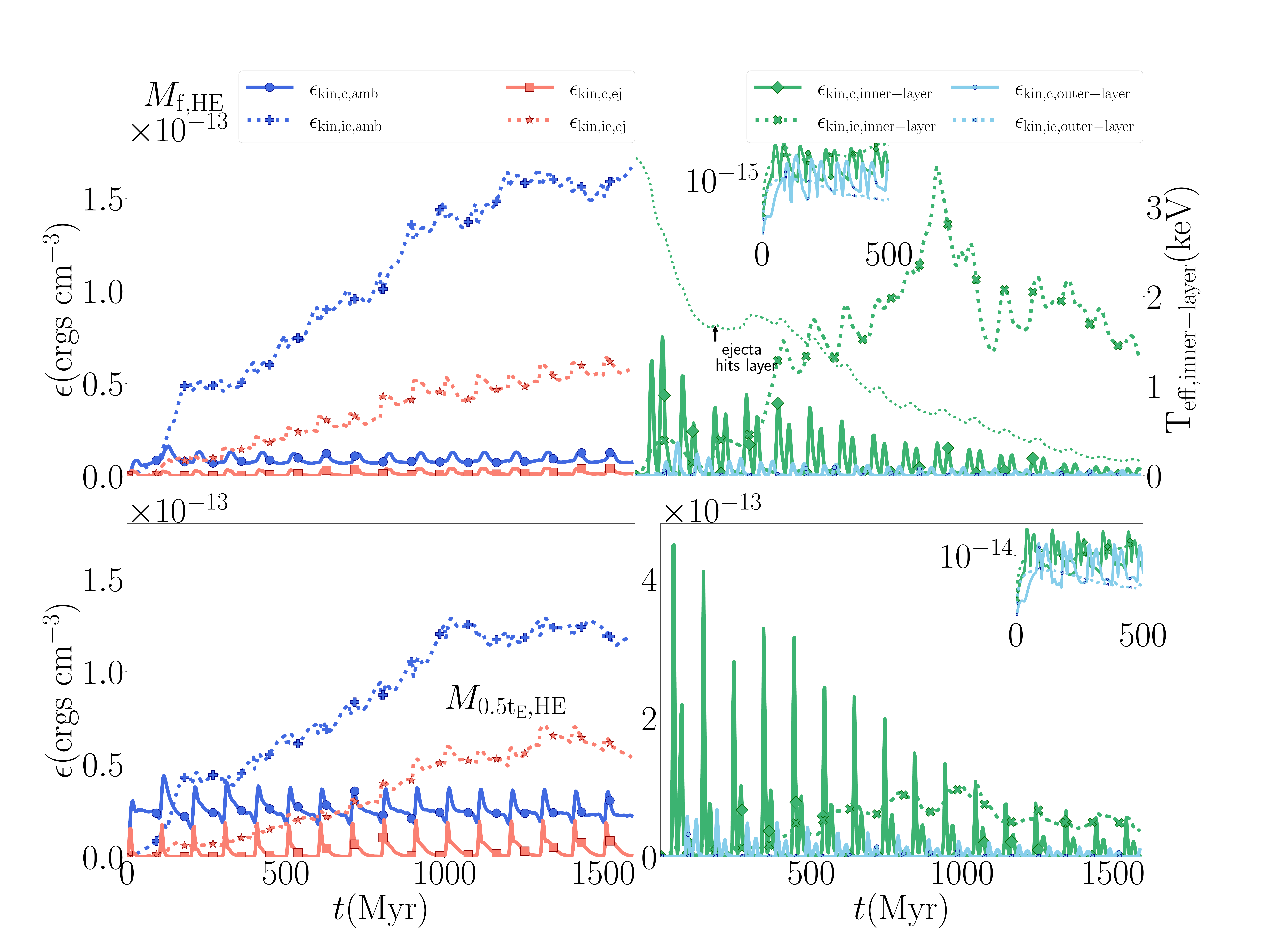}
    \centering
    \caption{The kinetic energy density (normalization by volume is important since in some cases we compare energy integrated over the inner uniform cube of the simulation box with that integrated within narrow layers within the box). The solid lines represent compressible mode ${\epsilon}_{\rm kin, c} = \int \rho u^2_{\rm c} dV/\int dV$ and dashed lines represent incompressible mode ${\epsilon}_{\rm kin, ic} = \int \rho u^2_{\rm ic} dV/\int dV$ (but see section \ref{sec: KEdensity} for details of normalization). Ejecta kinetic energy and ambient kinetic energy are normalized by the full box volume. The kinetic energies of the inner and outer dyed layers are normalized by the respective initial dyed volumes. The inset shows the similar early time evolution of ${\epsilon}_{\rm kin, inner-layer}$ and ${\epsilon}_{\rm kin, outer-layer}$. ${\epsilon}_{\rm kin, ic, inner-layer}$ is amplified around $\sim 500$ Myr. The second y-axis on the right in upper panel shows the volume-averaged temperature of the tracer dominated inner-layer and the arrow points at the time when the ejecta hits this layer for the first time.}
 \label{fig:KEdensity}
\end{figure*}   

\subsubsection{Can shorter $t_{\rm duty}$ and absence of cooling enhance sound wave energy?}
We include two additional cases in Figure \ref{fig:KE} which deviate from the trends of kinetic energy evolution in other simulations, namely, $L_{\rm 0.5t_E, 50t_d}$ and $L_{\rm 0.5t_E, nocool}$ (light brown and gray lines). In the former, the frequency of weak-shock injection is twice that of the fiducial cases and in latter we shut off the radiative cooling. The incompressible kinetic energy is reduced at all times in the former compared to all other cases and the sound wave energy is slightly elevated. Although the cooling flow is mildly delayed in this case (due to frequent injection of energy at the same rate as $L_{\rm 0.5t_E}$), we do not gain significantly in the sound wave energy. Moreover, the turbulent kinetic energy is reduced by almost a factor of $2$. Note that turbulence in itself can act as a source of sound waves (eq. 75.9 in \citealt{LANDAU1987251}).

The second case is interesting because it reduces both compressible and incompressible kinetic energy at all times. It means that sound waves are significantly amplified in presence of radiative cooling. Since the sound-crossing times in typical clusters are much shorter than the cooling times and in addition isentropic thermal instability is a rare occurrence (eq 5 in \citealt{1965ApJ_field}), it is unlikely that sound waves are thermodynamically affected by radiative cooling. Denser structures within $\sim 50$ kpc formed by inhomogeneous cooling possibly modify the sound waves by repeated reflections. Clearly, the gray wave-like pattern of the compressible energy in time in Figure \ref{fig:KE}, implies that in other cases the sound waves are modulated. 

It is worth mentioning that g-modes or the Brunt-V\"{a}is\"{a}l\"{a} oscillations may generate vortical motions. These are thermally unstable when the perturbations in the ICM (which is in thermal and hydrostatic equilibrium) are linear (\citealt{2016MNRAS_choudhury}). Hence g-modes can be amplified if the feedback is optimum. This cannot occur in the absence of cooling.

\subsubsection{Triggering g-modes by ejecta and transfer of excitation across radial layers: can turbulence fill up the volume?}
\label{sec: KEdensity}
The thermal energy that is injected in the $R_{\rm inj}$ kpc sphere causes the region to inflate. That converts a fraction of the energy to the kinetic form. This excitation can show up in both compressible and incompressible motions. Thus we see sound waves propagating out and local Brunt-V\"{a}is\"{a}l\"{a} oscillations (and/or hydrodynamic instabilities) being triggered. Within $\sim 50$ kpc, the latter is dominant.

In order to understand this process better, we consider $M_{\rm f, HE}$ and $M_{\rm 0.5t_E, HE}$ (see Table \ref{tab:sims}) which deposits more feedback energy. The purpose of injecting somewhat higher energy (and power) is also to explore what prevents cooling flow globally. In these simulations, we inject two types of tracers, ${\rm Tr}_{\rm layer}$ and ${\rm Tr}_{\rm ejecta}$ (in Figure \ref{fig:dye} these are called `Layer-dye' and `Ejecta-dye' respectively). The `Layer-dye' is injected at $50 (\pm 2)$ kpc (inner layer) and $100 (\pm 2)$ kpc (outer layer). The `Ejecta-dye' is injected with only the feedback energy which means it is replenished. Figure \ref{fig:dye} represents a volume-rendering at $3$ Myr and $1.48$ Gyr in $M_{\rm f, HE}$. This simulation does not show a cooling flow till $\sim 2.5$ Gyr (simulation ends) and the central $\sim 10$ kpc is slightly sparse (half the central density at $t=0$) at all late times. The volume rendering of the `Ejecta-dye' on the right shows a clear yellow zone which holds the largest fraction of the thermal energy injected. The ring of `Layer-dye' in the inner layer is completely disrupted but a significant portion of this matter seems to be inside $50$ kpc. The outer ring, while is wobbling, retains most of its content. The wobbled outer layer is clearly visible in the XY-plane of the inset plot. Overall the vorticity is not dominant beyond $50-60$ kpc. The central core in $M_{\rm 0.5t_E, HE}$ is very similar to $M_{\rm f, HE}$.

The tiny wobbling in the outskirts is not sustained by the initial density fluctuations we insert since those disappear in $\sim 100$ Myr. In order to get a quantitative understanding, Figure \ref{fig:KEdensity} shows the compressible and incompressible kinetic energy density (kinetic energy normalized by volume within which it is relevant). The blue, red and the green/cyan lines correspond to the matter with and without ejecta in the following way:
\begin{eqnarray}
\label{eq:KEdensity_all}
\nonumber
{\epsilon}_{\rm kin, amb} = \frac{\int \rho (1. - {\rm Tr}_{\rm ejecta}) u^2 dV_{\rm box}}{\int dV_{\rm box}}\\
\nonumber
{\epsilon}_{\rm kin, ej} = \frac{\int \rho {\rm Tr}_{\rm ejecta} u^2 dV_{\rm box}}{\int dV_{\rm box}}\\
{\epsilon}_{\rm kin, inner/outer-layer} = \frac{\int \rho {\rm Tr}_{\rm layer} u^2 dV_{\rm layer}}{\int dV_{\rm layer}}
\end{eqnarray}

Firstly, if we focus on the blue and red lines, the ambient and ejecta kinetic energies, we see that in the entire inner box (each of side $100$ kpc), the ambient medium surely dominates. This is visually evident from Figure \ref{fig:dye} since ejecta dyed volume is actually small compared to the entire box. However the ejecta may still excite the layers above. Also, note that the incompressible kinetic energy density dominates in the box. The green lines which almost follow the blue, correspond to ${\rm Tr}_{\rm inner-layer}$. Hence the ambient medium is dominated by layers of gas similar to what the green lines represent ($\sim 50$ kpc). However, the cyan lines, corresponding to the outer dyed layer, contribute little to the energy density budget. This verifies turbulence generated by the ejecta, is confined to inner layers. Beyond $\sim 400-500$ Myr, the inner layer energies rise steeply. That coincides with the time when ejecta first hits this layer. The fall of the green lines correspond to the de-localization of the inner layer's tracer material. Both $M_{\rm f, HE}$ and $M_{\rm 0.5t_E, HE}$ run till $2.5$ Gyr and the region beyond $\sim 50$ kpc never show strong vorticity. Large scale g-modes triggered beyond $\sim 50$ kpc, simply break into the central turbulent core. Thus there is always a strong confinement. 

To understand the higher peaks of kinetic energy of the inner layer for the initial $500$ Myr compared to those of the outer, we plot the effective temperature (the secondary y-axis in Figure \ref{fig:KEdensity}) of this inner layer as following: 
\begin{eqnarray}
\label{eq:KEdensity_T}
\nonumber
T_{\rm eff, inner-layer} = \frac{\int T {\rm Tr}_{\rm layer} dV_{\rm layer}}{\int dV_{\rm layer}}
\end{eqnarray}
The green dotted line in the Figure \ref{fig:KEdensity} (upper panel) shows this temperature with time. A black arrow shows the time of first hit. Before that the temperature falls steeply. After the first collision of the ejecta and the inner-layer, the ejecta starts hitting this layer frequently. The layer is gradually disrupted and the material is dispersed in the surroundings. 

In the lower panel ($M_{\rm 0.5t_E, HE}$), while we see broadly similar incompressible energy densities for the ambient medium, layers and the ejecta, there is an increased sound flux for all these components by a factor of $\sim 4$ (which roughly corresponds to the increase in instantaneous power in Table \ref{tab:sims}). 

\subsection{Where do we expect thermalization?}
\label{sec:whenwhere}
\begin{figure*}
    \includegraphics[width=\linewidth]{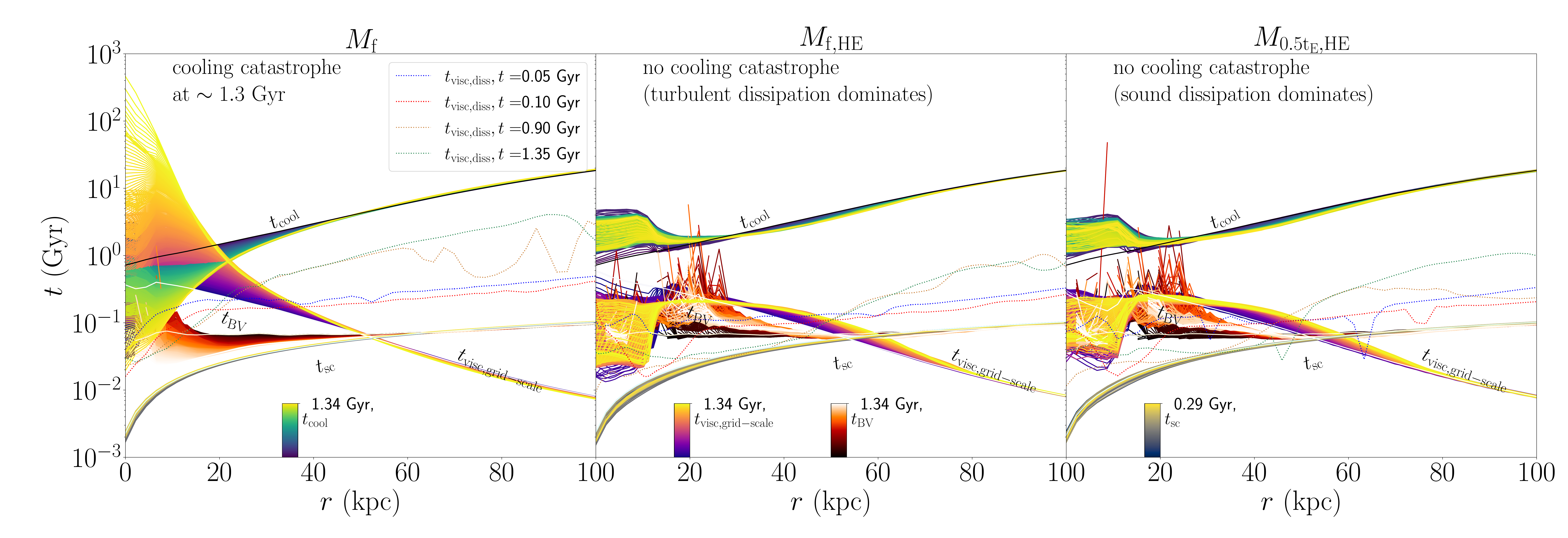}
    \centering
    \caption{The variation in the timescales, $t_{\rm cool}, t_{\rm BV} (1/N_{\rm BV}), t_{\rm visc}, t_{\rm visc, diss}$, of various physical processes across the time of evolution of the ICM. The time of evolution changes according to the colormaps at the bottom of the plot, except for $t_{\rm visc, diss}$, which is labelled on the legend of the left column. The solid black, white and cornsilk lines are the $t=0$ profiles of $t_{\rm cool}$, $t_{\rm visc}$ and $t_{\rm BV}$. Note that the outgoing sound waves are expected to be fast and not dissipate within $50$ kpc. The g-modes that contribute to the central turbulence may dissipate within $\sim t_{\rm visc, diss}$ and heat the central gas. The central dip and plateau in $t_{\rm visc}$ and $t_{\rm cool}$ respectively in $M_{\rm f, HE}$ are set by the ejecta domain (while $R_{\rm inj}/R_{\rm E}$ is same for all cases, $R_{\rm inj}$ is larger in $M_{\rm f, HE}$ and $M_{\rm 0.5t_E, HE}$). The background sound crossing time is not as variable as the other timescales, so plotted over a shorter range as shown in the rightmost colorbar.}
 \label{fig:timescales}
\end{figure*}   

\begin{figure*}
    \includegraphics[width=18cm]{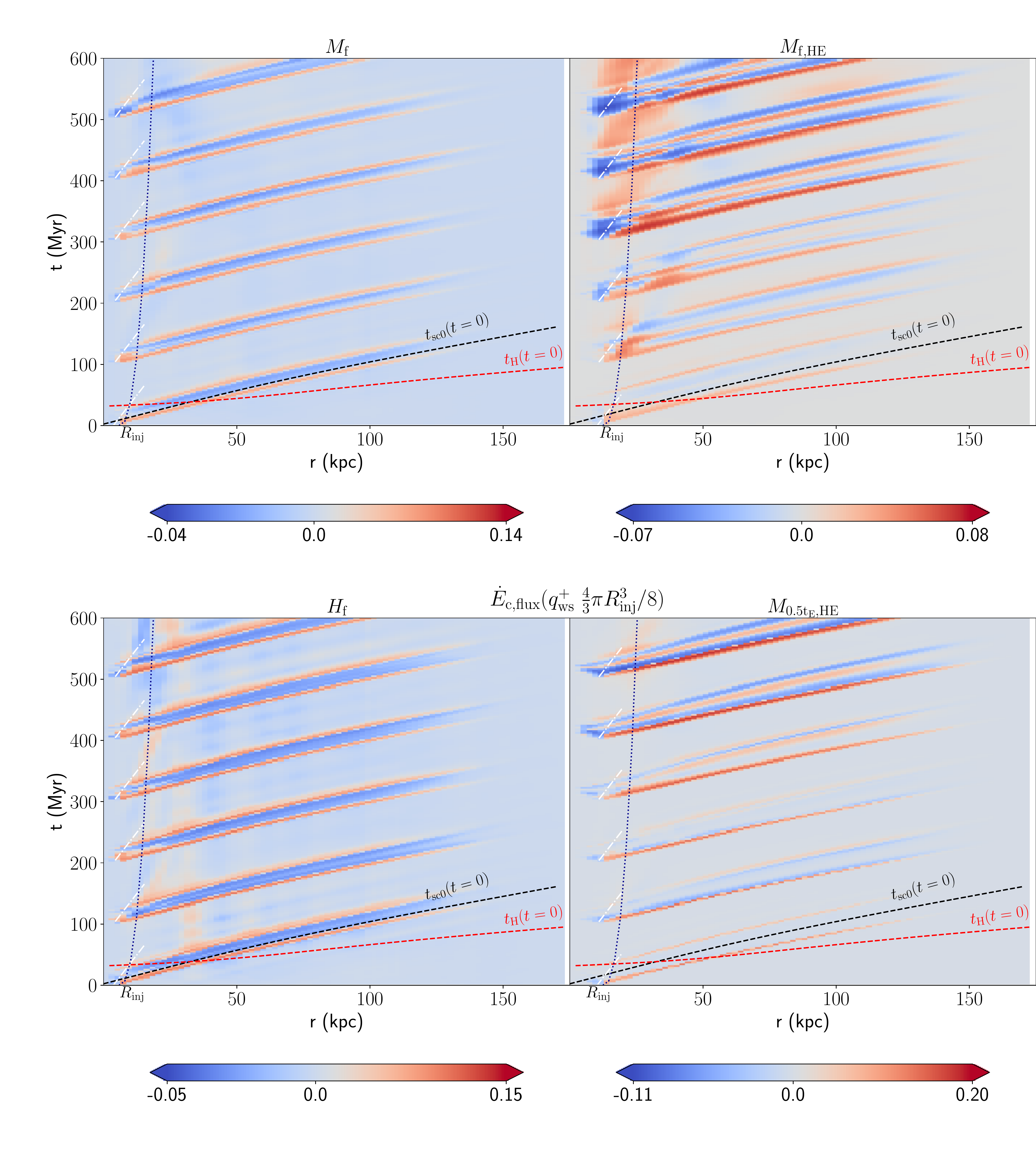}
    \centering
    \caption{Normalized flux (flux calculated in octants as in eq \ref{eq:soundflux}) of compressible disturbance across each of $N_{\rm r}$ shells in $M_{\rm f}$ ($N_{\rm r}=80$), $M_{\rm f, HE}$ ($N_{\rm r}=80$), $H_{\rm f}$ ($N_{\rm r}=80$, here we keep physical width same; see section \ref{sec:diag}), and $M_{\rm 0.5t_E, HE}$ ($N_{\rm r}=80$). Note that the normalization constant is not necessarily same (see Table \ref{tab:sims} for injection energy and radius,  $E_{\rm inj}$ and $R_{\rm inj}$). The black dashed lines show the sound crossing time in the hydrostatic background ICM and the white dashed lines represent a disturbance that propagates at $5$ times slower speed than sound ($c_{\rm s0}/5$). The white dashed lines are drawn from $R_{\rm inj}$ at every $t_{\rm spread}$ within the initial $500$ Myr. The dark blue dotted line shows a pure self-similar expansion of the ejecta starting from $t=0$. The red dashed lines show the time taken to cross a pressure-scale height $H=\frac{c^2_{\rm s0}}{\gamma g}$ locally. Sound waves may be attenuated above the radius of intersection of the black and red dashed lines since $t_{\rm H}$ is shorter.}
 \label{fig:sflux}
\end{figure*}

\begin{figure*}
    \includegraphics[width=18cm]{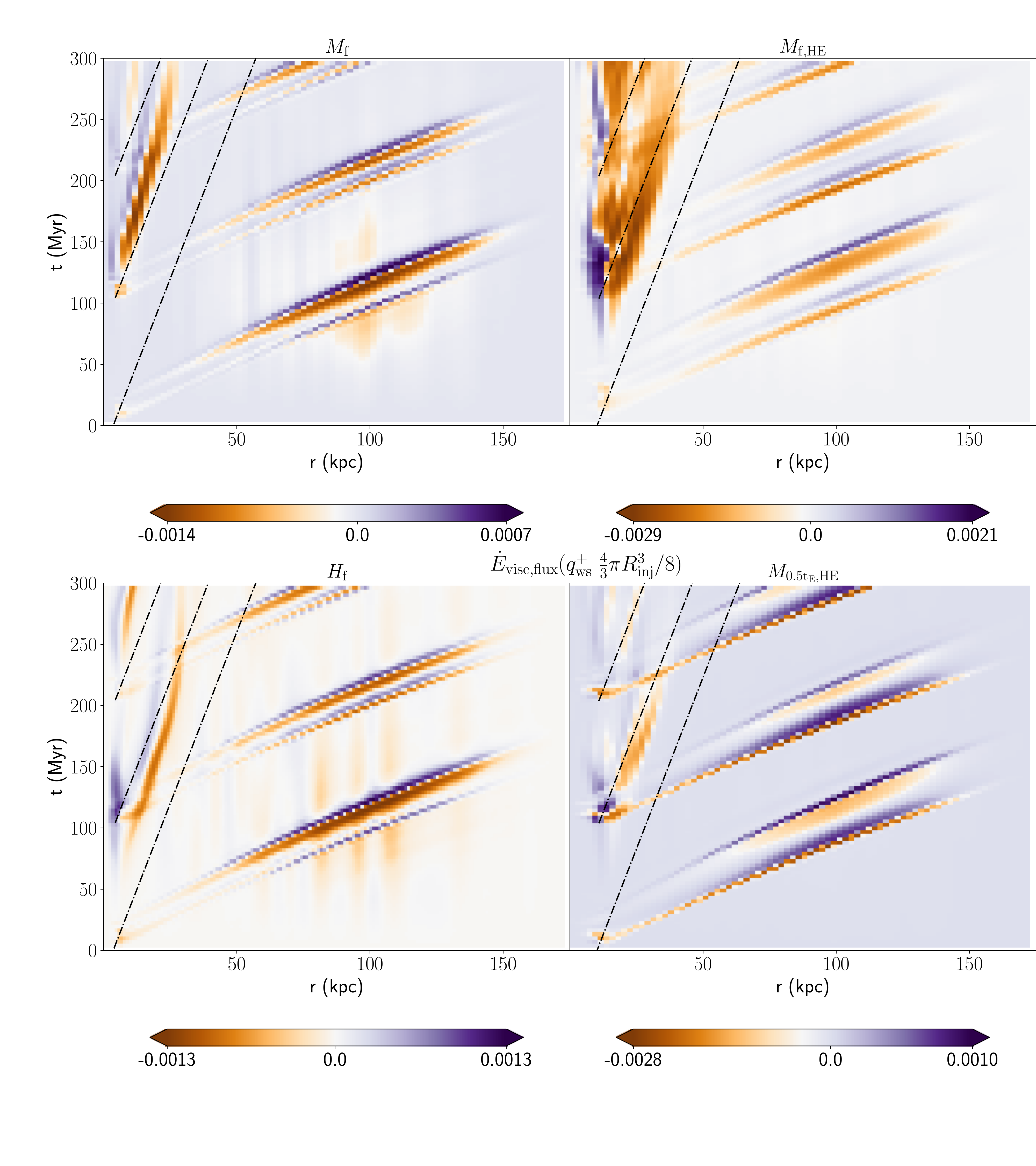}
    \centering
    \caption{Viscous flux (eq \ref{eq:viscflux}) for $M_{\rm f}(N_{\rm r} = 80)$, $M_{\rm f, HE}(N_{\rm r} = 80)$, $H_{\rm f}(N_{\rm r} = 160)$, and $M_{\rm 0.5t_E, HE}(N_{\rm r} = 80)$, including the trajectories of subsonic expansion (black dashed) as in Figure \ref{fig:sflux}. The striped patterns from lower left to upper right coincide with the sound trajectories. Viscous flux is extremely high within and around the ejecta radius (along the dashed lines) as soon as the timescale is comparable to  $t_{\rm visc, grid-scale}$ and $t_{\rm visc, diss}$. }
 \label{fig:vflux}
\end{figure*}

\begin{figure*}
    \includegraphics[width=18cm]{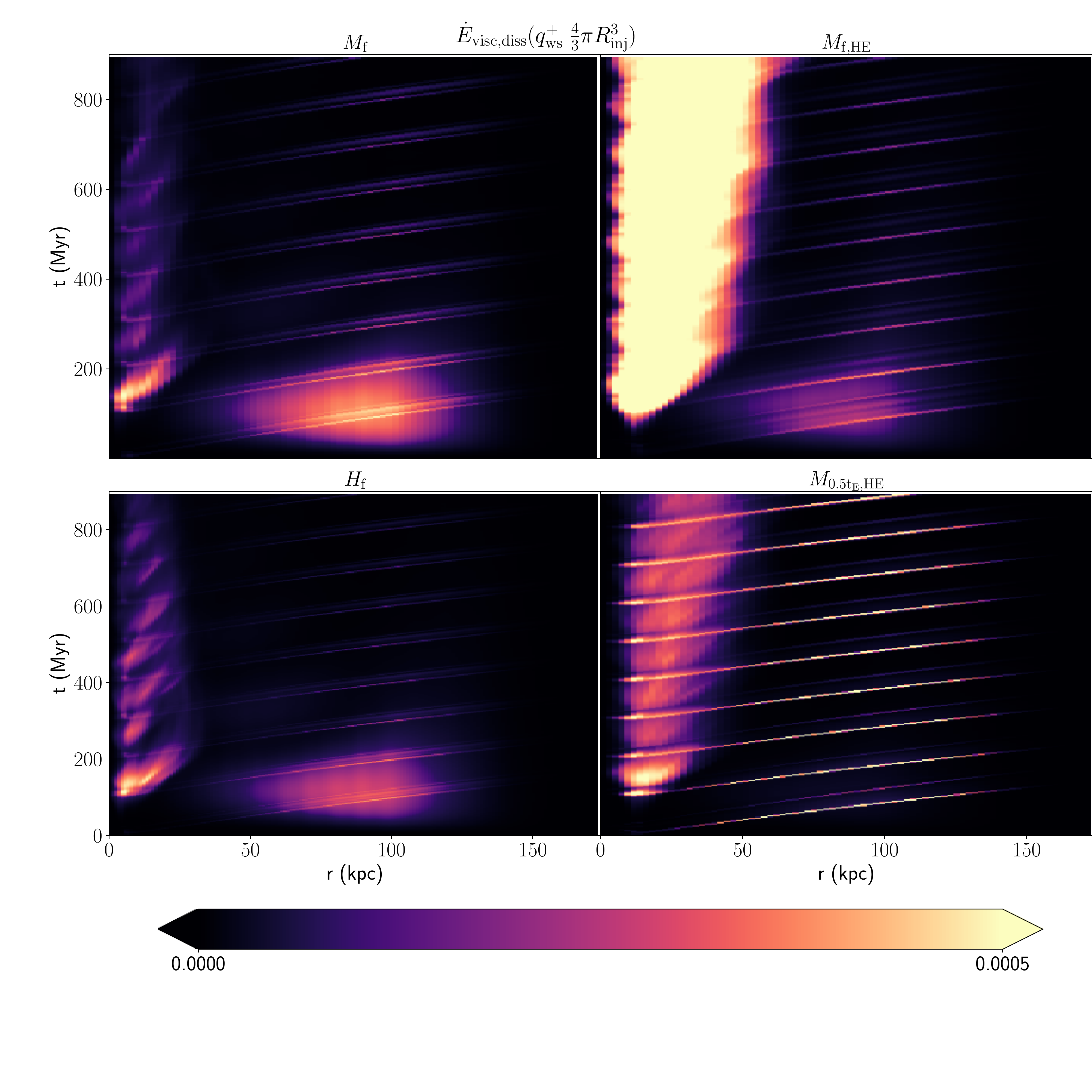}
    \centering
    \caption{Viscous dissipation rate (eq \ref{eq:viscdiss}) for $M_{\rm f}(N_{\rm r} = 80)$, $M_{\rm f, HE}(N_{\rm r} = 80)$, $H_{\rm f}(N_{\rm r} = 160)$, and $M_{\rm 0.5t_E, HE}(N_{\rm r} = 80)$ for $900$ Myr. The striped patterns from lower left to upper right coincide with the sound and viscous flux trajectories.  Note the maximum dissipation rates in $M_{\rm f, HE}$ and $M_{\rm 0.5t_E, HE}$ on the right panels are greater than that of left panels by a factor of $\lesssim 10$.  We fix the colormap to spatially locate the large dissipation rates in different simulations.}
 \label{fig:viscdiss}
\end{figure*}

\begin{figure*}
    \includegraphics[width=18cm]{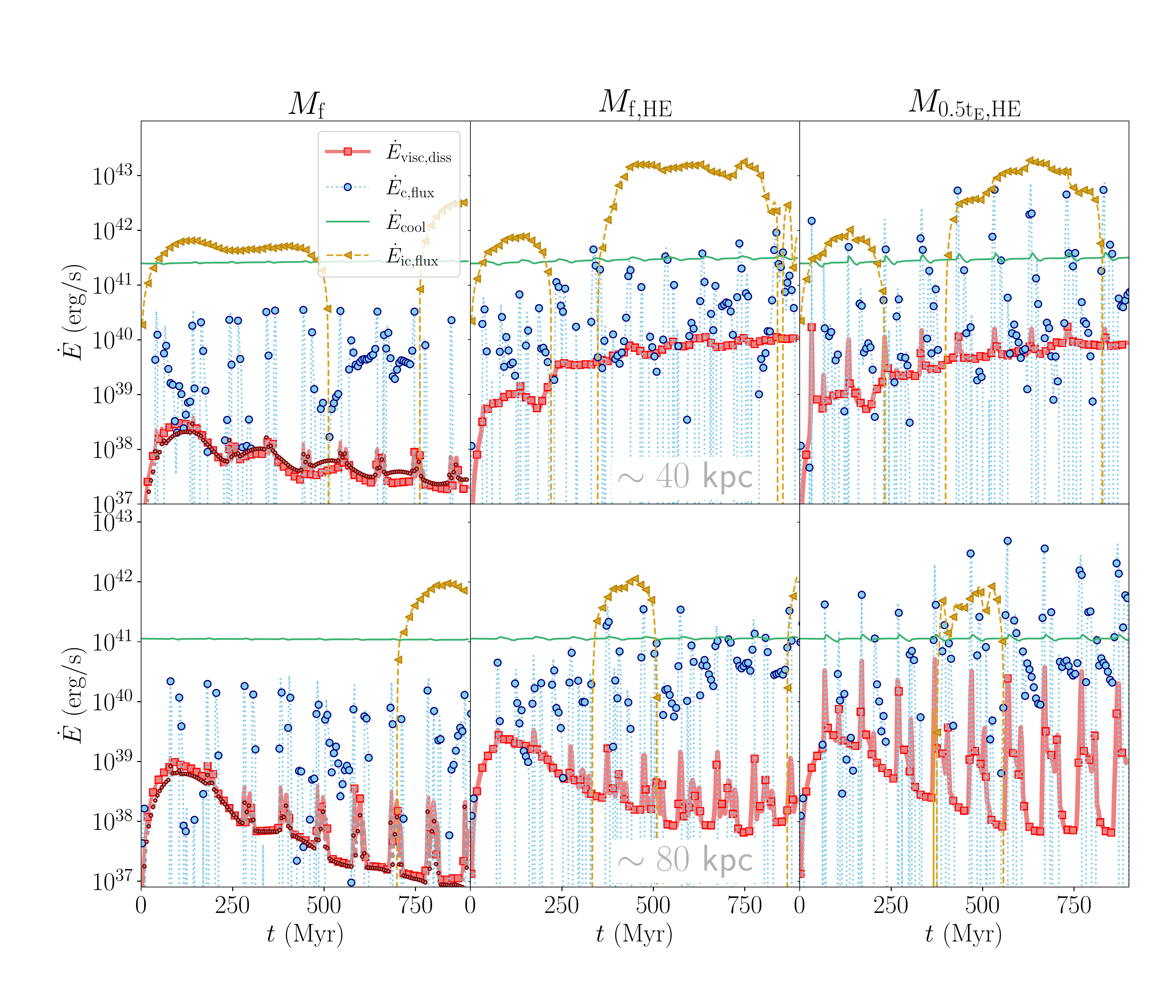}
    \centering
    \caption{The compressible and incompressible fluxes (in the octant of maximum compressible flux), cooling and dissipation rates (averaged over octants) across specific radial shells (total $80$ shells within a sphere of radius $100$ kpc but including the corners of the cubic box that encloses it, hence $\triangle r \approx 2$ kpc). Each frame is equivalent to a vertical slice of the corresponding cases in Figure \ref{fig:sflux}, \ref{fig:vflux} and \ref{fig:viscdiss}. Note that on the left column we show points from $H_{\rm f}$ in $\dot{E}_{\rm visc, diss}$ up to $\sim 1$ Gyr on top of the red markers to highlight the level of convergence between simulations of different resolutions. It indicates that enhanced dissipation by progressively smaller scales will be dominant only within the central ejecta zone (discussed more in section \ref{sec:lnu}). For a brief discussion on the dynamic effect of viscosity in the simulations using same diagnostics, see Figure \ref{fig:diss_extend}.}
 \label{fig:edot}
\end{figure*}
From our analysis in section \ref{sec:v_comp_incomp}, we understand that there are complex dynamics associated with compressible wave-like motion (sound waves) and vorticity (g-modes and hydrodynamic instabilities). In addition, as cooling starts dominating, there is pressure gradient driven motion in the center. Sound waves of moderate wavelengths should propagate out very fast while we find that turbulence is more confined. The viscous dissipation may occur due to thermalization of sound waves or the turbulence - both may contribute to heating. In order to assess the role of g-modes and sound waves in heating, we need to understand the timescales of thermalization of these waves. 

Figure \ref{fig:timescales} shows the relevant timescales for $M_{\rm f}$, $M_{\rm f, HE}$, and $M_{\rm 0.5t_E, HE}$ varying across space and time. Firstly, the four colormaps show the variables plotted and the color value corresponds to the evolution time. $M_{\rm f, HE}$ (middle) and $M_{\rm 0.5t_E, HE}$ (right) are quite similar. In $M_{\rm f}$, the cooling time becomes very short in the center at latter times (yellow), while in $M_{\rm f, HE}$(\& $M_{\rm 0.5t_E, HE}$), it is maintained closer to the initial values. If the core does not collapse, region outside $70-80$ kpc is not expected to cool ($t_{\rm cool} \sim 10$ Gyr). That is exactly what happens in the `no cooling catastrophe' cases. $t_{\rm BV}$ is quite short between $10-40$ kpc in $M_{\rm f}$ (white), while in $M_{\rm f, HE}$(\& $M_{\rm 0.5t_E, HE}$), $t_{\rm BV}$ at late times oscillate around the initial cornsilk colored solid line. This reflects that the gas entropy gradient is large for $M_{\rm f}$ between $10-40$ kpc (we ignore the inner $\sim 10$ kpc for both cases for fair comparison, since the ejecta thermalizes predominantly there), and the buoyancy oscillations are faster than radiative cooling process. Fast oscillations can barely take part in the global dynamics. The dissipation timescale between $10-40$ kpc progressively becomes higher than $t_{\rm BV}$ (green and yellow dotted lines). On the contrary in $M_{\rm f, HE}$ between $10-40$ kpc, slower buoyancy oscillations cause two simultaneous conditions: (i) some regions may be thermally unstable and form cooler gas out to this radius, or in other words, may form multiphase regions (yellow and orange spikes reaching the $t_{\rm cool}$), (ii) trigger a more sustained, long-term turbulence due to g-modes which may dissipate fast (note the reverse evolution of $t_{\rm BV}$ and the $t_{\rm visc, diss}$ compared to $M_{\rm f}$).  

Sound waves are expected to be only mildly affected by the central dynamics in the inner $50$ kpc since these propagate out fast (although turbulence can source additional sound waves). As discussed in sections \ref{sec:timescales} and \ref{sec:th_sw}, linear sound waves (small-amplitude) decay due to viscous forces in a timescale $\sim \nu k^2$ while the incompressible velocities may dissipate in $\sim t_{\rm visc, diss}$ (eq \ref{eq:vdiss_t} in section \ref{sec:timescales}). At grid scales (smallest scale in our simulations but $> l_{\nu}$ mentioned in section \ref{sec:th_sw}), the decay timescale becomes smaller than the sound crossing time ($t_{\rm sc}$) beyond $\sim 60$ kpc. Roughly, we may expect the sound waves to thermalize outside $\gtrsim 50$ kpc. On the other hand, turbulent motions, if any, dissipate in the ``outskirts" (beyond $\gtrsim 60$ kpc) at a longer timescales (dashed lines) relative to that in the center. This difference between inner region and outskirts is less for $M_{\rm f, HE}$. Also, the buoyancy oscillation timescales are closer to $t_{\rm visc, diss}$ in $M_{\rm f, HE}$ in the region $10-40$ kpc. Hence the g-modes driven turbulence will dissipate fast in the central region. 

\subsubsection{How are $M_{\rm f, HE}$ and $M_{\rm 0.5t_E, HE}$ different?}
Notably, the middle and right panel look very similar since both reflect overall thermally stable cluster cores. We will discuss in the following sections on how only one of the energy carriers (sound or turbulence) dominates the heating in each case. Here we only point to the expected scenarios according to Figure \ref{fig:timescales}. Turbulence can heat the central $\sim 50$ kpc while sound waves can heat beyond $\gtrsim 50$ kpc. If there is a strong sound flux (linearly we expect long-wavelength waves to travel larger distances and there will always be loss of short-wavelength waves closer to the source), these waves must dissipate fast in the outskirts in $\lesssim 0.1$ Gyr while turbulence always dissipates over timescales longer than sound crossing time in the center. For weaker sound flux, waves are possibly easily scattered off the central inhomogeneities and the turbulence dominantly heats the center.

\begin{figure*}
    \includegraphics[width=16cm]{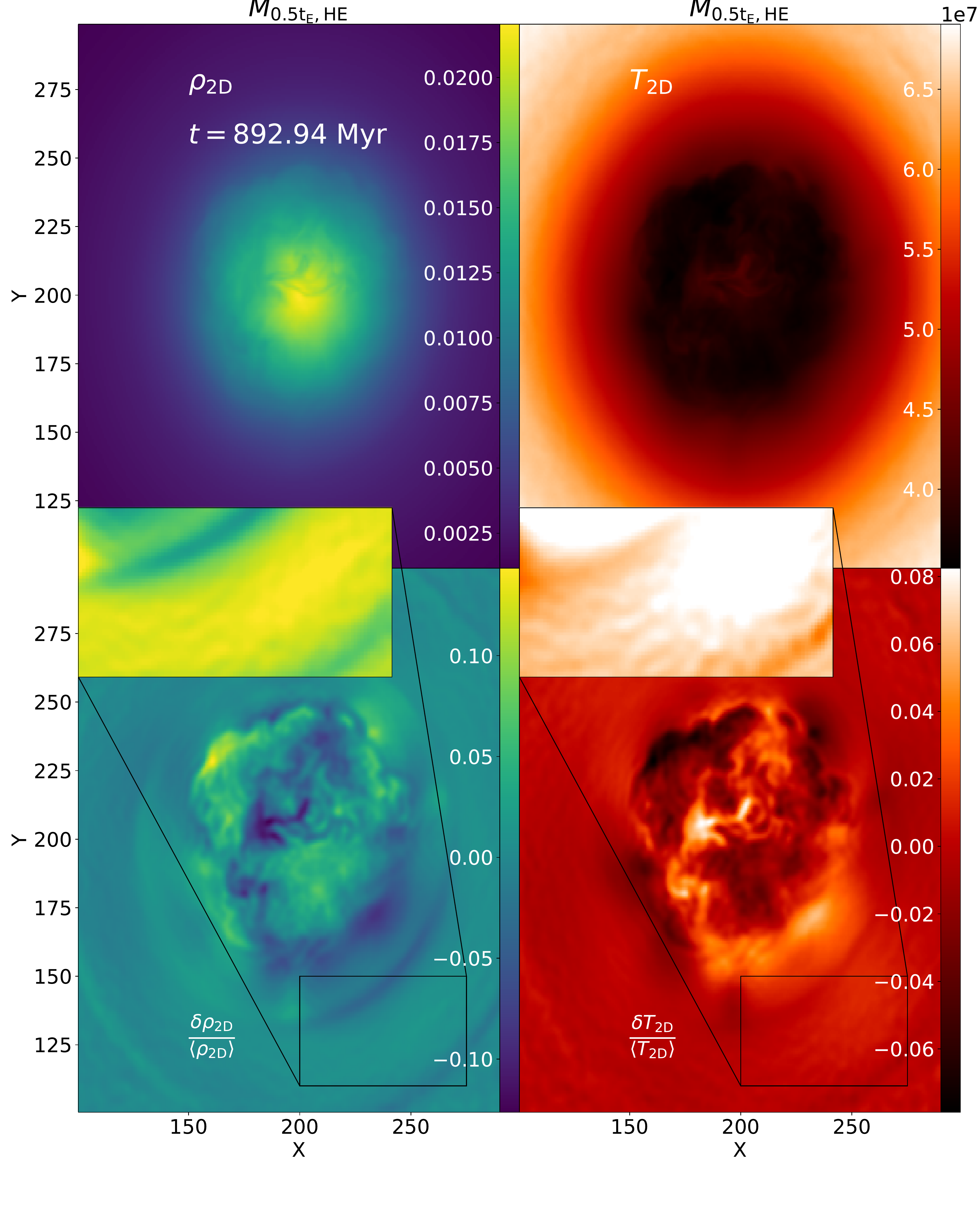}
    \centering
    \caption{The upper two panels show emission weighted density and temperature integrated along z-axis for a given time (eq \ref{eq:emm2d}). The bottom panels show the fluctuations in ${\rho}_{\rm 2D}$ and $T_{\rm 2D}$ by subtracting radial means ${\langle \rho \rangle} (r)$, ${\langle T \rangle} (r)$. Small patches of the maps are zoomed in to highlight sound wave ripples. In the zoomed panels we modify the colomaps to visualize smaller contrasts.}
 \label{fig:obs0}
\end{figure*}
\begin{figure*}
    \includegraphics[width=18cm]{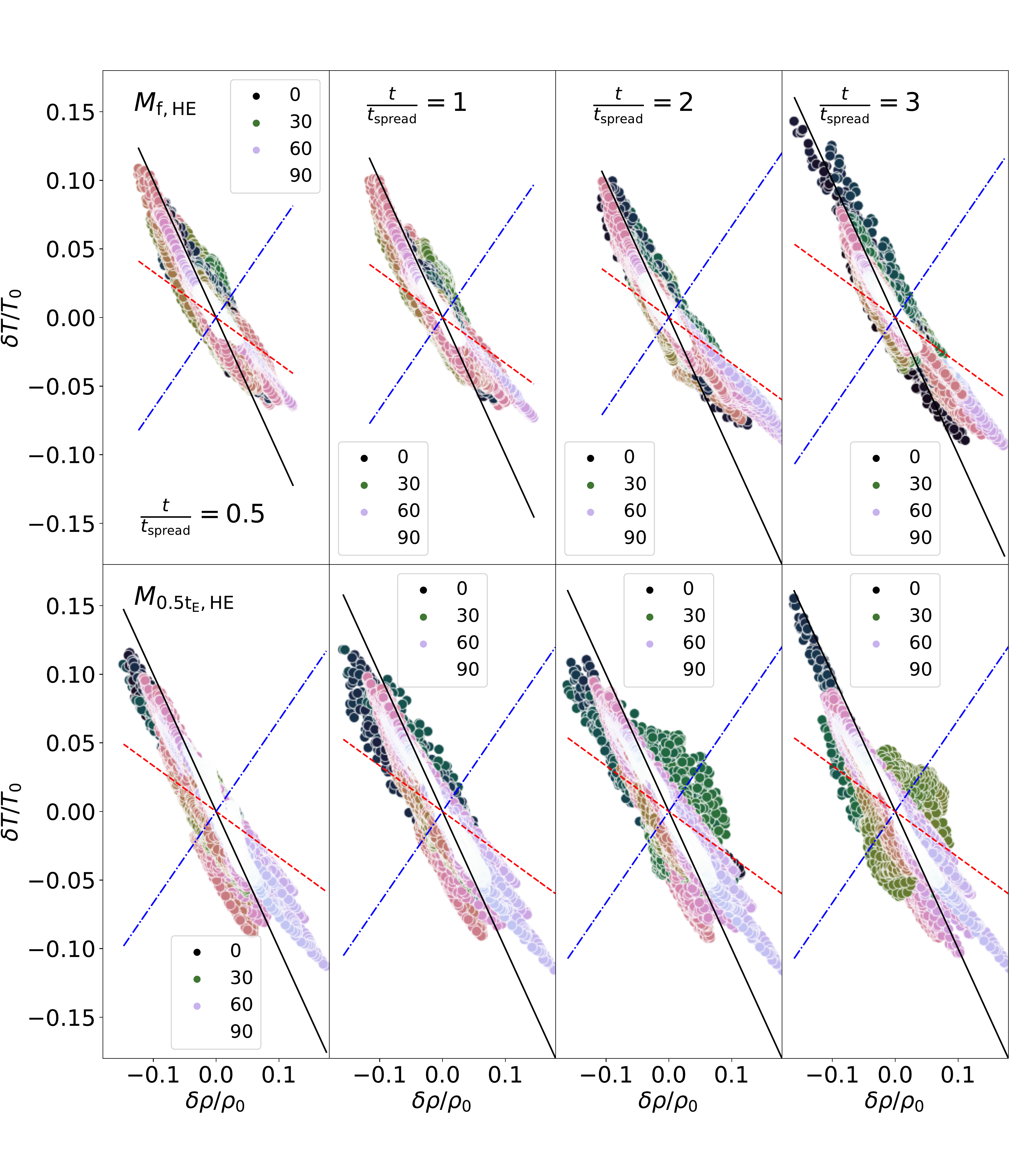}
    \centering
    \caption{The correlation between the fluctuations in emission weighted density and temperature in one cycle of feedback starting at $1.5$ Gyr.The colormap of the markers represent the radial distance from the center (black: closest, white: farthest). The black solid line corresponds to ${\Big[\frac{\delta T}{T_0}}\Big]_{\rm isob} = -\frac{\delta \rho}{{\rho}_0}$. The blue dash-dotted line corresponds to ${\Big[\frac{\delta T}{T_0}}\Big]_{\rm adiab} = (\gamma -1)\frac{\delta \rho}{{\rho}_0}$ and the red dashed line shows ${\Big[\frac{\delta T}{T_0}}\Big]_{\rm tot} = {\Big[\frac{\delta T}{T_0}}\Big]_{\rm isob} + {\Big[\frac{\delta T}{T_0}}\Big]_{\rm adiab}$ as a reference line.}
 \label{fig:obs2}
\end{figure*}
\subsection{Propagating sound waves}
\label{sec:swsim}
In this section, we will focus on the propagation of sound waves. The first $300-500$ Myr has a significant overlap between all the interesting physical timescales as shown in Figure \ref{fig:timescales}.

In order to understand the propagation, we calculate the sound flux across the surface of spherical shells considered in the inner uniform box of side $200$ kpc (the largest sphere in it has a diameter $200$ kpc). In order to filter compressible disturbance, we calculate the compressible velocity fluctuations in the inner uniform simulation box at each cell. The compressible flux is given by \footnote{In this method of shell-averaging or calculating flux on a shell surface, we note that in the central region the resolution is poorer compared to outer shells since the number of cells in cartesian geometry is less for smaller shells. On the other hand, beyond $r=100$ kpc the measurement is based on a relatively smaller number of cells than inner shells since these regions are outside the largest sphere that fits into the box.},
\begin{eqnarray}
\label{eq:soundflux}
\dot{E}_{\rm c, flux} = {\int}^{\rm r + \triangle r}_{\rm r} \delta p  ~\mathbf{u_{\rm c}} \cdot d \mathbf{S_{\rm cart}}
\end{eqnarray}
where $\delta p = p - \langle p \rangle (r)$ is the pressure fluctuation. The flux is calculated on the individual shells in each octant of the sphere and the octant with the highest flux is considered. The flux is not large in magnitude in all directions. The flux may decay (in magnitude) more in directions where gas cools sufficiently closer to the center.  

Figure \ref{fig:sflux} shows the normalized flux in the octant of highest flux for four important cases from Table \ref{tab:sims}. We plot the sound crossing time of the background ICM in black dashed lines, and the white dashed lines show the trajectory of a subsonic ($c_{\rm s0}/5$) expansion that approximately follows the pattern on the upper left corner of the plot (it roughly follows the evolution of $R_{\rm inj}$). We have also included evolution (in dark blue dotted lines) of a pure self-similar expansion, starting from $t=0$ at $R_{\rm inj,0}$. Theoretically, each blast event is expected to evolve by both free expansion followed by self-similar evolution. But we have a complex core with multiple ejections and the blue line simply implies that each ejecta region radially extends quite slowly (but note in $H_{\rm f}$ in Figure \ref{fig:vflux}, with thinner shells, we capture signatures of self-similar evolution in the upper left corner for the second outburst). In all four cases, we can see the series of sound waves propagating out along the theoretical trajectory (black dashed) after being generated on the surface of $R_{\rm inj}$. The lows imply either the trough of the leading wave or a reflection. The maximum flux is in $M_{\rm 0.5t_E, HE}$ which is $\lesssim 20$\% of the instantaneous power. $M_{\rm f, HE}$ generates minimum forward flux since a large fraction of the compressible flux is confined around the ejecta. We note here that in $M_{\rm f}$ and $H_{\rm f}$, a cooling catastrophe is delayed only by $\sim 400$ Myr ($\sim 1.3$ Gyr; without feedback, the catastrophe is over by $900$ Myr) while in $M_{\rm f, HE}$ and $M_{\rm 0.5t_E, HE}$, there is no signature of a catastrophe till $\sim 2.5$ Gyr. Despite a relatively large fraction of compressible flux, adequate feedback energy $E_{\rm inj}$ is necessary to prevent catastrophe. Hence $M_{\rm f}$ and $H_{\rm f}$ fail.

\subsection{Dissipation of the waves and turbulence}
In order to understand how turbulence and sound waves contribute to the heating of the ICM, we calculate the viscous flux ($\dot{E}_{\rm visc, flux}$ within the same octants of highest compressible flux) and the viscous dissipation rate ($\dot{E}_{\rm visc, diss}$ including all directions) and focus on the spatio-temporal evolution of these in this section. 
\begin{eqnarray}
\label{eq:viscflux}
\dot{E}_{\rm visc, flux} = {\int}^{\rm r + \triangle r}_{\rm r} \mathbf{\Pi}\cdot \mathbf{u} \cdot d \mathbf{S_{\rm cart}}
\end{eqnarray}
\begin{eqnarray}
\label{eq:viscdiss}
\dot{E}_{\rm visc, diss} = {\int}^{\rm r + \triangle r}_{\rm r} {\varepsilon}_{\rm diss} dV
\end{eqnarray}
where ${\varepsilon}_{\rm diss}$ is given by eq \ref{eq:vdiss}. Figures \ref{fig:vflux} and \ref{fig:viscdiss} show the viscous flux and the viscous dissipation rate. The negative viscous flux in $M_{\rm f, HE}$ is prominent in the core region while in other cases the viscous flux is typically higher in along the sound trajectories and the ejecta radius (slanted upper left patterns). The latter happens since the ejecta region is expanding slightly.  


The viscous dissipation in Figure \ref{fig:viscdiss} (note that the maximum in the right panels is greater than that on the left panels by an order of magnitude but we fix the colormap to assess the relative difference in spatial heating) is consistent with the viscous flux.  From Figure \ref{fig:sflux}, we know that the number of sound trajectories (marked by positive flux) is larger in $M_{\rm f, HE}$. But the flux amplitude is smaller. It is possible that sound waves disperse in the medium and different frequencies interfere with each other well inside the central region. This leads to a loss of sound flux and more dissipation within the inner $50$ kpc. Consequently in $M_{\rm f, HE}$ both compressible and incompressible motions dissipate reasonably fast in the core and never allow it to collapse. In $M_{\rm f}$ and $H_{\rm f}$, there is no sustained dissipation but only the dissipation of compressible modes that dominate along (and within) the ejecta boundaries. While $M_{\rm f, HE}$ seems a prospective feedback scenario, $M_{\rm 0.5t_E, HE}$ shows another viable and efficient mechanism to distribute heat even outside $\gtrsim 50$ kpc by sound waves. The strongest sound wave flux in this simulation leads to a significant heating rate of the outskirts in the wake of the propagating waves (bright yellow stripes in the lower right panel). The larger the sound flux, the faster will be the decay rate in the outskirts since timescale depends on the local temperature for a given wavelength. On the other hand, visually the turbulence (or vorticity) map is very similar in the cores of $M_{\rm f, HE}$ and $M_{\rm 0.5t_E, HE}$. However, the net turbulent dissipation rate in the core is also dependent on the background temperature; formation of ejecta with even mildly lower enthalpy may cause $\nu$ (and hence dissipation rate) to be somewhat lower in $M_{\rm 0.5t_E, HE}$ core. But compressible dissipation dominates even within the ejecta since the compressible velocities (hence $\delta u$) are larger than in $M_{\rm f, HE}$. 

\subsection{Prevention of a cooling catastrophe}
In this section, we will discuss whether exact cooling-heating balance is necessary to preserve thermal stability in the ICM. In order to understand this energetics, we consider a radial location ($\sim 40$ kpc and $\sim 80$ kpc ) in the r-t plots of previous section and assess the fluxes, cooling rate and dissipation rate there.
The dissipation rate and compressible flux (note that this flux in an octant is indicative of the maximum impact sound waves may have on the outskirts) are same as shown in eqs \ref{eq:soundflux} and \ref{eq:viscdiss}. The cooling rate is the following:
\begin{eqnarray}
\label{eq:clngrate}
\dot{E}_{\rm cool} = {\int}^{\rm r + \triangle r}_{\rm r} n_{\rm e} n_{\rm i} \Lambda (T) dV
\end{eqnarray}

The incompressible radial flux going out (or in) across this radius is,
\begin{eqnarray}
\label{eq:incompflux}
\dot{E}_{\rm ic, flux} = {\int}^{\rm r + \triangle r}_{\rm r} p \mathbf{u}_{\rm ic} \cdot d\mathbf{S}_{\rm cart}
\end{eqnarray}
where $p$ is the local pressure and other symbols have usual meanings. As usual, this flux is calculated only in the octant of maximum sound flux. To compare physical values of the fluxes and energy rates better, we take a simple octant average ($/8$) of the cooling/heating rates. Hence, we note that fluxes indicate an upper limit. 

In Figure \ref{fig:edot} we show the $\dot{E}$ in cgs unit only in the positive axis (fluxes can be negative). If we compare the plots in the upper panel between $M_{\rm f}$ and $M_{\rm f, HE}$, the primary difference is the presence of a continuous dissipation rate in time rather than spiked dissipation rate coinciding with the spikes of sound transit (red thick lines). A crucially interesting feature is also that the net dissipation increases with time in $M_{\rm f, HE}$. However both the characteristics appear in $M_{\rm 0.5t_E, HE}$, spiked compressible dissipation and gradually increasing turbulent dissipation. Generally, the incompressible flux (yellow) is relatively more dominant at small radii which is consistent with a central turbulent regime but for $M_{\rm f}$ that is clearly not a sufficient condition to prevent subsequent cooling catastrophe. Compressible flux (blue) is the largest in $M_{\rm 0.5t_E, HE}$. In the lower panels (at $\sim 80$ kpc) we again note that a combined effect of heating and large outgoing sound flux to be effective. Note that the first bump of dissipation rate (also present in the upper panel for $M_{\rm f}$) is due to velocity fluctuations induced by initial small density perturbations. Further, note the maroon line in the left most panels represent the dissipation rate from the $H_{\rm f}$ simulation. The red and maroon lines imply that there is a good convergence. In the upper-left panel there is higher dissipation rate in $H_{\rm f}$  at late times since smaller grid scales may generate dissipative contribution from waves/fluctuations of smaller length-scales in the ejecta region.

The $\dot{E}_{\rm cool}$ (green line) in all the four plots show tiny perturbations with each sound transit event. This is extremely important since such fluctuations are observed in X-ray surface brightness and are discussed extensively. Note that $M_{\rm f, HE}$, which is preventing cooling catastrophe more efficiently, has somewhat stronger indication of sound passage in $\dot{E}_{\rm cool}$ (larger and extended spikes in the green line). The strongest spikes occur in $M_{\rm 0.5t_E, HE}$. Hence observations of ``sound" fluctuations in the surface brightness will surely help to assess if sound waves are indeed playing an interesting role in global energetics. We will discuss how $M_{\rm 0.5t_E, HE}$ may appear in observations in the next section.

\subsection{What do we expect in X-ray observations?}
\label{sec:surfb}
In this section, we will discuss how our simulated ICM may appear in X-ray surface brightness image. This will help to determine the underlying physical process for a given feature in such an image. 

Firstly, we calculate the emission-weighted density ($\rho$) and temperature ($T$) maps integrated along the line-of-sight (along z):
\begin{eqnarray}
\label{eq:emm2d}
{\rho}_{\rm 2D} = {\int}^{z_{\rm max}}_{z_{\rm min}} \rho n_{\rm e} n_{\rm i} \Lambda(T) dz/{\int}^{z_{\rm max}}_{z_{\rm min}} n_{\rm e} n_{\rm i} \Lambda(T) dz \\
T_{\rm 2D} = {\int}^{z_{\rm max}}_{z_{\rm min}} T n_{\rm e} n_{\rm i} \Lambda(T) dz/{\int}^{z_{\rm max}}_{z_{\rm min}} n_{\rm e} n_{\rm i} \Lambda(T) dz
\end{eqnarray}
As mentioned in section \ref{sec:diag} for 3D, the fluctuations in ${\rho}_{\rm 2D}$ and $T_{\rm 2D}$ in the 2D plane perpendicular to the line-of-sight, can be calculated by using an area-averaged radial mean in each radial shell ($\langle {\rho}_{\rm 2D} \rangle $ and $\langle {T}_{\rm 2D} \rangle$), then subtracting the radial mean from the density and temperature fields respectively. Since the central dynamics is governed by a highly disturbed, strong vorticity-filled inner core, we will investigate the outer regions closely. Note that a simulation like ours is an ideal testbed to assess how to extract information about sound waves from observed data, since we are injecting sound waves manually into these clusters. We know from our fluxes that sound waves are present out to $100$ kpc. This helps to conclude what happens to the signals of sound waves in emission-weighted maps.

Figure \ref{fig:obs0} shows the ${\rho}_{\rm 2D}$ and $T_{\rm 2D}$ maps in the upper panels for a given time in $M_{\rm 0.5t_E, HE}$. As discussed earlier, the central $50$ kpc is appropriately captured to be highly turbulent in these maps and compared to that inner core, a fairly quiet exterior. The lower panel shows the fluctuations after subtracting $\langle {\rho}_{\rm 2D} \rangle$ and $\langle {T}_{\rm 2D} \rangle$ respectively. Visual inspection suggests that the central region is completely dominated by isobaric fluctuations with precise anti-correlation between $\delta T/T_0$ and $\delta \rho/{\rho}_0$ (Note that we refer to the backgrounds generally with subscript `0'). We consider a rectangular patch away from the center and re-plot the fluctuations here with a modified colormap (lower maximum and higher minimum) to capture finer contrasts. The moderately long wavelength sound waves that propagate from the source (surface of $R_{\rm ej})$, maintain quasi-linearity and dissipate in the outskirts. The zoomed-in panels in Figure \ref{fig:obs0} clearly have striped patterns in fluctuations which appear to be along the circumference of a distorted circle centered at the ICM core. The positive correlation between $\delta T/T_0$ and $\delta \rho/{\rho}_0$ is visible along these fluctuations.  

\subsubsection{Correlation between $\delta T/T_0$ and $\delta \rho/ {\rho}_0$}
In order to understand the thermodynamic state of the fluctuations in the ICM, or more precisely, how the fluctuations may appear in the observations, it is useful to assess the correlation between $\delta T/T_0$ and $\delta \rho/ {\rho}_0$ quantitatively. 

Figure \ref{fig:obs2} shows the $\delta T/T_0$ versus $\delta \rho/ {\rho}_0$ trend for four time frames within a given outburst cycle. We consider a cycle at $1.5$ Gyr which is well into the regime in which the ICM is affected by repeated outbursts already. The colors of the markers represent the distance from the center. In the upper panel ($M_{\rm f, HE}$), all the points collapse within a narrow range around isobaric black line while in the lower panel ($M_{\rm 0.5t_E, HE}$), there is a spread along the adiabatic line (blue dash-dotted) at later times. The spread reflects on how the fluctuations look as a transient wave transits. Typically the spread along the adiabatic line has similar color at any point in time which implies it is a very localized scenario as expected. Over time the fluctuations drop back to the black isobaric line until the next outburst. Since the sound flux is weaker in $M_{\rm f, HE}$, the sound passage is not as strongly reflected in a spread along the blue line. 

There is another interesting point to note: almost all the fluctuations tend to align with the black line after the transit of the sound, but at the large $\delta \rho/ {\rho}_0$, the fluctuations move away from the black line towards larger $\delta T/T_0$. This implies that there is slight heating in those fluctuations. On the other end, the black points align closer to the black line at most times since at small radii isobaric perturbations are dominant. Nevertheless a slight cooling (deviation below black line) is visible. To summarize, in X-ray observations when we assess the equation of state from fluctuations, we may find a strong alignment along the black line, missing out the adiabatic spread since sound waves are transient and localized at any given time. It is useful to understand these subtle features of slight heating/cooling and the `adiabatic spread' despite a strong isobaric alignment.

\section{Discussions}
\label{sec: disc}
We carry out a suite of three-dimensional hydrodynamic simulations of viscous ICM to understand the dynamics of sound wave and energy transport through it in the context of the convectively stable galaxy clusters. The source of the energy in the ICM is AGN feedback triggered by the microphysical accretion process which we do not model here. We inject thermal energy $E_{\rm inj}$ for a duration of $t_{\rm spread}$ every $t_{\rm duty}$ as an emerging feedback scenario. Our goal is to understand how this energy can be gently circulated in and around the core. Contrary to powerful feedback (\citealt{2022arXiv_wang}), we focus on more gentle scenarios. Thus we avoid generating extremely strong shocks. A large fraction of the thermal energy gets converted to compressible and incompressible kinetic energy. Incompressible kinetic energy (turbulence) is always larger than the compressible kinetic energy (weak shocks/waves) but their roles in heating are reversed depending on the outburst duration. For a very slow rate of injection ($t_{\rm spread}/t_{\rm E}=2$), incompressible energy dominates the dissipation and this heating occurs preferentially in the central core. For a faster rate of injection ($t_{\rm spread}/t_{\rm E}=0.5$; slower than powerful jets), compressible flux carries away as high as $\lesssim 21\%$ of the power to large distances and dissipates dominantly in its wake. We further note than exact balance in heating and cooling may not be required to prevent a catastrophe since the outgoing flux can also support the core. In X-ray observations, fluctuations derived from projected surface brightness may appear to be dominantly isobaric turbulence (with large amplitudes) despite the important role of compressible waves. Adiabatic fluctuations are possibly detectable as spatially localized small amplitude features with $\delta T/T_0 = (\gamma -1) \delta \rho/{\rho}_0$. In the following sections we discuss several interesting implications of all our results summarized above. 

\begin{figure}
    \includegraphics[width=9cm]{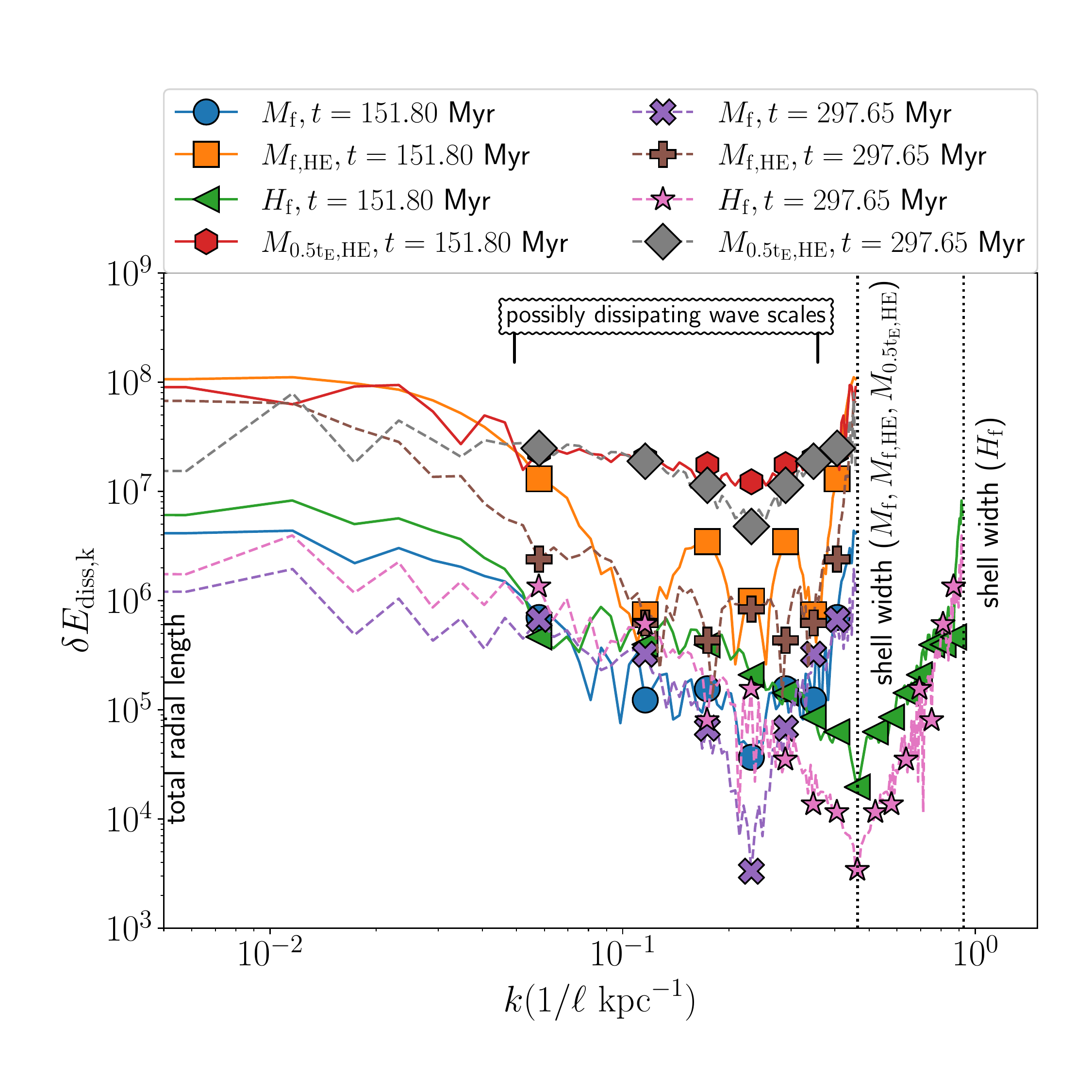}
    \centering
    \caption{The fourier transform of the $\dot{E}_{\rm visc, diss}$ (in code units) for two different times selected from Figure \ref{fig:viscdiss}. The earlier time ($151.8$ Myr) is also an earlier phase after a fresh injection while the later time ($297.65$ Myr) is a very late phase before a fresh injection. The black dotted lines on the right show the smallest scale ($1/dr$) in the radial grid we consider in medium and high resolution simulations. The lower limit on the x-axis corresponds to the total radial length scale ($1/N_{\rm r} dr$).}
 \label{fig:kspace}
\end{figure}
\subsection{Is it necessary to resolve $l_{\nu}$ to understand viscous dissipation that we see in the outer core?}
\label{sec:lnu}
It is ideal if we can resolve all the important smallest length/timescales to conclude about different physical scenarios in the ICM. That is not necessarily within the scope of current computational resources since $l_{\nu}$ may vary across orders of magnitude. Here we speculate why such unattainable resolution may not be needed to capture the long-wavelength ($\gtrsim$ kpc) sound wave dissipation in $M_{\rm 0.5t_E, HE}$ specifically. 

Our $t_{\rm visc}$ is simply a guiding timescale to estimate the approximate location of sound dissipation. Dissipation of sound waves dominantly occurs at the location where decay time of sound waves $\sim 1/\nu {k^2}_{\rm grid} <= t_{\rm sc}$. This can be re-written as $t_{\rm visc, grid} \sim dt_{\rm sc} \implies l_{\rm grid} \sim \frac{\nu}{c_{\rm s}} = 2 l_{\nu}$. But our grid scale in full simulations (${256}^3$:$0.78$ kpc, ${512}^3$:$0.39$ kpc) is larger than $l_{\nu}$ calculated from the ICM background by a factor of $>>2$ (background $l_{\nu}$ is at $\sim 10-100$ pc length scale in the outskirts). Even if we consider reduced velocity of linear sound waves in presence of viscosity (eq \ref{eq:omegnostr}), we get an estimate of length scale $\sim 2.2 l_{\nu}$. Despite not resolving $l_{\nu}$, in Figure \ref{fig:viscdiss} we see that the locations of viscous dissipation in the outskirts are identical in $M_{\rm f}$ and $H_{\rm f}$. Small-scale (high-$k$) sound waves are anyway susceptible to viscous decay spatially close to center (this will occur closer and closer to the center with increasing resolution). Hence what we see in the outer core is probably the dissipation of larger scale waves (note an approximate estimate of the length scale of such linear sound waves is given in section \ref{sec:sw_cool} for hot, dense gas). The linear sound waves can be attenuated gradually beyond $40-50$ kpc (see red dashed lines in Figure \ref{fig:sflux}). In addition, the answer to this may also be non-linear steepening of the long wavelength modes resulting in distortion and dissipation beyond the core. With better resolution, naively we might assume that $t_{\rm visc, grid}$ decreases and hence dissipation of sound waves must occur increasingly closer to the core, which doesn't happen starkly when we compare the two resolutions. Hence we predict that we already capture intermediate scale sound waves fairly well.

In order to test the resolution effects we plot the 1D Fourier transform of $\dot{E}_{\rm visc, diss}$ in the $k (1/\ell)$ space in Figure \ref{fig:kspace} to understand at which length scales energy dissipation increases for different simulations. Note that this plot is only indicative since the actual grid scale is not the smallest scale here; the radial shell width used in analysis is the smallest length scale. We still capture resolution effect since for larger resolution, we use proportionally larger number of radial shells in analysis. Firstly, we note that the lines for $M_{\rm f}$ and $H_{\rm f}$ (green-blue, violet-pink) are very close with latter decreasing slowly in the large-$k$ direction. This implies that the dissipation rate at intermediate scales in $H_{\rm f}$ may be higher (yet less than maximum). On the other hand, for $M_{\rm 0.5t_E, HE}$ the rates (red and gray) in this intermediate range is much higher physically. Hence the sound dissipation must be happening in these scales. An equivalent $H_{\rm 0.5t_E, HE}$ must show comparable dissipation rates in this range (given our $M_{\rm f}$ and $H_{\rm f}$ trend). Overall with higher resolution, we may get only mildly larger dissipation rates due to availability of smaller length scales in large-$k$ direction. Finally the increase in dissipation at the smallest available scale in all the simulations can be due to both distorted sound waves or turbulence in the core. 

\subsection{Is it possible to reconcile the wide range of sound escape fractions in earlier simulations?}
\citealt{2017MNRAS_tang} has classified AGN feedback based on outburst durations but the simulation is carried out in a simple spherical geometry. This precludes the g-mode dynamics and multiphase nature of the core. Both can enhance the sound wave fraction. \citealt{2019ApJ_bambic} and \citealt{2022arXiv_wang} simulate the most powerful outbursts and produce strong shocks in the core. While the former work does not distinguish between compressible waves and shocks, the latter find that compressible shocks carry a dominant fraction of the energy rather than the waves. In our work (particularly $M_{\rm 0.5t_E, HE}$), we avoid strong shocks and possibly this intermediate regime is where we can acquire maximum compressible energy yet not strong enough shocks that dissipate all the energy in the center. Hence despite having viscous effects dynamically included in our simulation, we get the propagating sound waves to carry a large flux out in $M_{\rm 0.5t_E, HE}$.
\subsection{Turbulence or sound waves?}
 Sound waves and turbulence are fundamental entities in any astrophysical fluids. But there are disjoint ideas regarding the role of these two fundamental energy carriers in feedback. From X-ray surface brightness observations, earlier works have extracted density fluctuations (see our section \ref{sec:surfb} for comparable calculations done with our idealized simulation). The analysis based on the surface brightness fluctuations highlight that dominant fluctuations in the medium are isobaric (e.g., \citealt{2018ApJ_zhuravleva}) and hence must be due to turbulence. {\tt Hitomi} used soft Xray spectra to obtain moderate gas velocities within $60$ kpc. But further recent works (\citealt{2019NatAs_zh}) claim that ICM is possibly very weakly viscous and hence numerical models with high Reynolds number (another characteristics of turbulence) are appropriate. Suppressed viscosity is in fact theoretically justified in presence of magnetic fields. However, none of these disprove the presence and role of adiabatic sound waves in contributing to feedback. None of the previous explorations can also precisely point to the exact roles of either turbulence or sound waves in heating the core or outskirts. 

In our simulations, we see that the turbulence is dissipated fast enough in the central region in our $M_{\rm f, HE}$ (successful in preventing cooling flows). We also see another successful scenario in $M_{\rm 0.5t_E, HE}$, in which sound waves carry away energy from the center to dissipate in the outskirts while the core is dominated by swirling turbulent motions. Thus both the scenarios are likely depending upon the feedback duration. These results are not only consistent with observations but also help to reconcile the past results. We also find a more realistic escape fraction of sound ($\lesssim 20$ \%) in 3D simulations as opposed to 1D/2D done earlier.

\subsection{Convective stability/instability, anisotropic transport with magnetic fields}
In this work, we model a convectively stable ICM. For a hydrodynamic cluster, the stability criterion depends on the gas entropy gradient and the latter becomes crucial for the evolution of waves and saturation of any instability. However, in presence of small magnetic fields, thermal conduction occurs along field lines and the convective stability criterion depends on the temperature gradient; if temperature and pressure gradients are along the same direction, there is magnetothermal instability (MTI) and if those are in opposite directions there is heat-flux buoyancy-driven instability (HBI) both of which may trigger vigorous turbulence. The instabilities saturate at suppressed heat flux reorienting magnetic fields (\citealt{2007ApJ_parrish, 2008ApJ_parrish}). Note that anisotropic viscosity may impair such reorientation and help in conduction (\citealt{2012ApJ_kunz}). But on the other hand, kinetic particle-in-cell simulations highlight that thermal conduction regulated by whistler waves inhibits itself (\citealt{2016ApJ_roberg-clark, 2018JPlPh_komarov}). Hence it is unclear if convection in turn is diminished enough to consider hydrodynamic cluster to be a good representation of reality. We can only speculate that the resultant impact could be short phases of stronger turbulence due to convective motions.

How acoustic propagation evolves in presence of micro-instabilities is still a nascent area to explore. Non-linear saturation of fast magnetosonic waves is often evoked to account for sonic turbulence in high $\beta$ (thermal to magnetic pressure ratio) plasma. Earlier works claimed that Alfv\'enic fluctuations mix phases of compressive fluctuations but recent analytic work shows that for sufficiently large $\beta$, it doesn't hold since kinetic instabilities are generated which prevent phase mixing. The consequence is that sound waves of reasonable amplitude escape before being damped in weakly collisional plasma (\citealt{kunz_squire_schekochihin_quataert_2020}). This restores our hydrodynamic picture as the emergent phenomenon. But it is also imperative to explore the details of sound propagation in presence of magnetic turbulence and instabilities in future. 

\subsection{Implications of cosmic rays for sound waves}
It has been recently revealed by \citealt{2020MNRAS_kempski} that sound waves or the fast magnetosonic waves are linearly overstable in presence of cosmic rays (CRs). They argue (in section 3.6) that long wavelength ($\sim$ few kpc) sound waves are more susceptible to large amplification and least affected by kinetic microinstabilities. They also point out that for sufficiently large $\beta$ the growth rate of the sound waves may amplify density perturbations by $10\%$ by the time the waves propagate out to $50$ kpc in $50$ Myr (in our simulations sound waves propagate approximately similar distance). This implies that in localized regions, sound waves may appear in the large amplitude fluctuations of surface brightness (or ripples) as has been claimed to be seen in Perseus (\citealt{2006MNRAS_fabian}). In our $M_{\rm 0.5t_E, HE}$, the adiabatic fluctuations are as high as $\lesssim 10$ \% and with CRs these can be further enhanced.

\section{Conclusions}
We use idealized hydrodynamic simulations of the intracluster medium to study how thermal energy derived from moderately powered AGN feedback is gently circulated within and outside the core. We draw the following conclusions:
\begin{itemize}
    \item {\bf Compressible \& solenoidal energy fractions: }Total magnitude of incompressible (g-modes/instabilities) kinetic energy at any given time for all simulations is higher than the compressible kinetic energy by a factor of $\lesssim 10$ (see Figure \ref{fig:KE} and the blue lines of ambient kinetic energy in Figure \ref{fig:KEdensity}). It is further interesting that in presence of radiative cooling, both the components are enhanced significantly. Internal gravity waves are known to be thermally unstable hence the increase in incompressible kinetic energy is justified. It is evident that compressible modes are also sourced by g-mode turbulence. There is no significant effect of varying initial injection radius ($R_{\rm inj}/R_{\rm E}$) on this result. Overall, our conclusion is that a large fraction of pure thermal energy injection at kpc scales always produces core-filling ($\lesssim 50$ kpc) swirling motions (turbulence).
    \item {\bf Role of escaping sound waves: }Propagating sound waves are abundant in our ICM. With each injection event, we find trains of sound waves traveling up to $100$ kpc and beyond. The net sound flux after repeated outbursts can go as high as $\lesssim 20$\% in a moderately powerful feedback. These become weaker in the outer core due to a combined impact of acoustic cut-off frequency, interaction with local linear g-modes, viscous losses and eventually dissipate non-linearly. Consequently if we can generate a very strong sound flux (in $M_{\rm 0.5t_E, HE}$), the long wavelength modes dissipate in the outskirts very efficiently. We conclude that these are the long wavelength sound waves since these travel out to a large distance to dissipate (not very dependent on resolution, see left column in Figure \ref{fig:viscdiss}). Short wavelength sound waves (high $k$) must decay closer to the center (at rate $\sim \nu k^2$ which must rise with resolution) but are possibly dominated over by incompressible dissipation (at grid scales possibly as a turbulent cascade) within inner core, specifically in $M_{\rm f, HE}$). 
    \item {\bf What's an efficient feedback in our simulations? } In our suite of simulations, $M_{\rm f}$ is inefficient while $M_{\rm f, HE}$ and $M_{\rm 0.5t_E, HE}$ are efficiently preventing cooling flows. Let us address the prevention issue by posing two questions: i) what is a sufficient feedback? ii) what happens in real clusters? The first question is answered by the $M_{\rm f, HE}$ scenario in which turbulent dissipation and possibly short-wavelength sound dissipation both occur in the inner $50$ kpc. A significant outgoing compressible flux is also built up and combined with the heating, it successfully stabilizes the cluster. The second question captures the problem of understanding observations and reconciling those with simulated clusters. We anticipate that $M_{\rm 0.5t_E, HE}$ shows another viable feedback circulation in which sound waves dominantly heats the cluster, specifically in the outskirts of core. A stronger outgoing sound flux is also contributing to energy balance and efficiently maintains the thermal equilibrium. This is also consistent with observed ripples in the outskirts of Perseus core that have been proposed to be sound waves.
    
    \item {\bf Where are all the sound wave fluctuations gone? } Despite finding the ICM to be infested with sound waves (more in $M_{\rm 0.5t_E, HE}$), if we calculate the emission-weighted density and temperature 2D maps and find the respective fluctuations, we see adiabatic fluctuations at low-to-moderate amplitudes intermittently. The sound perturbations are sub-dominant in the center. We see circular patterns in the outskirts at $\lesssim 10$ \% fluctuations. However, all the fluctuations are dominantly aligned to an isobaric state (Figure \ref{fig:obs2}) after the passage of sound. This reflects that in surface brightness fluctuations it could be difficult to track the sound waves and precisely determine their role. The spread of the isobaric fluctuations along the respective adiabats may indicate localized sound waves in a given radius. Minute deviations towards over-heating/cooling away from the isobaric state can also signify the passage of sound.  
\end{itemize}
\label{sec:cons}
%
%
\section*{Acknowledgements}
We acknowlege the support of Cambridge Service for Data-Driven Discovery (CSD3). All simulations have been run in the Peta4 supercomputer under CSD3. We are grateful to the frontline health workers all over the world who are fighting the ongoing COVID-19 crisis. PPC acknowledges Andy Fabian for useful discussions. PPC also acknowledges the talks and discussions during the KITP `halo21' program 2021 supported by NSF PHY-1748958. The authors further acknowledge the European Research Council (ERC) for support under the European Union\textquotesingle s Horizon 2020 research and innovation programme (project DISKtoHALO, 
grant 834203).

\section{Data Availability}
The relevant data will be made available on any reasonable request to the authors.

{\it Softwares used: } We have used mostly publicly available softwares and libraries in this work, namely, {\tt PLUTO}, numpy, scipy, matplotlib, cython, ParaView and Mathematica. Animations of two of the simulations are available in the following \href{https://youtube.com/playlist?list=PLNcUKdSsVbbElNxPt7Lpr2CZ1YPqQD9Mo}{link}.
%




\bibliographystyle{mnras}
\bibliography{bibtex} 

%
%
%
\appendix
\section{Linear coupled system of sound waves and g-modes}
\label{sec:applin}
In presence of a gravitational potential (along x), the gas layers are stratified and we need to take into account the gradients of the variables like density, entropy, etc. Such a system hosts g-modes which may or may not be coupled (linearly) to sound waves. Without loss of generality, we take the wave vector $\mathbf{k_{\rm R}}$ in the $xy$-plane. Here the fluctuations take a form $\delta F = F^{\prime} e^{i(kx + ly - \omega t)}$ where $l$ denotes the perpendicular wave-number and the other variables have usual meanings as in section \ref{sec:sw_simplest}. The linearised equations are similar to section \ref{sec:sw_simplest} with additional terms (we do not take Boussinesq approximation so that sound waves and g-modes can appear simultaneously but we only consider gradient of entropy to be significant and neglect other gradients),
\begin{eqnarray}
-i \omega {\rho}^{\prime} &=& - ik{\rho}_0 u_1^{\prime} - il{\rho}_0 u_2^{\prime}\\
-i \omega {\rho}_0 u_1^{\prime} &=& -ikp^{\prime} + ({i})^2 {\eta}_{\rm c} T^{\frac{5}{2}}_{70} [A_{1 \nu}u_1^{\prime} +B_{1 \nu}u_2^{\prime}] - g {\rho}^{\prime}\\
-i \omega {\rho}_0 u_2^{\prime} &=& -ilp^{\prime} + ({i})^2 {\eta}_{\rm c} T^{\frac{5}{2}}_{70} [A_{2 \nu}u_1^{\prime} +B_{2 \nu}u_2^{\prime}] \\
-i\omega s^{\prime}/s_0 &=& - \frac{\gamma N^2_{\rm BV}}{g} u_1^{\prime}
\end{eqnarray} 
Here $\eta_{\rm c} = \xi_{\nu} \mu m_{\rm p} {10}^{25}$, $s = \ln(p/{\rho}^{\gamma})$, $N_{\rm BV} = \sqrt{g \frac{ds}{dx}/\gamma}$ is the Brunt-V\"ais\"al\"a frequency (when medium is convectively stable and goes through buoyancy oscillations). Also, $s^{\prime}/s_0 = p^{\prime}/p_0 - \gamma {\rho}^{\prime}/{\rho}_0$ such that $p^{\prime} = \Big(p_0 s^{\prime}/s_0 + c_{\rm s0}^2 {\rho}^{\prime}\Big)$, $A_{1\nu} = \frac{4}{3}k^2 + kl$, $B_{1\nu} = k^2 - \frac{2}{3}kl$,$B_{2\nu} = \frac{4}{3}l^2 + kl$, $A_{2\nu} = l^2 - \frac{2}{3}kl$. Note that in the absence of viscous terms and filtering out sound waves (Boussinesq approximation), the dispersion relation of g-modes in a system like the above is ${\omega}^2 = \frac{l^2 N^2_{\rm BV}}{l^2 + k^2}$.

The dispersion relation for the coupled viscous sound waves and g-modes is given by:
\begin{eqnarray}
\label{eq:fulldisp}
\nonumber
{(i \omega)}^5 - {(i \omega)}^4 (2 \bar{\nu}B_{2\nu} +\bar{\nu}A_{1\nu}) &+& {(i \omega)}^3 \Big[{\bar{\nu}}^2(B^2_{2\nu}\\
\nonumber
+ 2 A_{1\nu}B_{2\nu} - A_{2\nu}B_{1\nu}) &+& c^2_{\rm s0}k^2_{\rm R} - ikg \Big(1+ \frac{c^2_{\rm s0}N^2_{\rm BV}}{g^2} \Big)\Big]\\
\nonumber
+ {(i \omega)}^2 \Big[ \frac{2ik c^2_{\rm s0} N^2_{\rm BV} \bar{\nu} B_{2\nu}}{g} &-& \frac{il c^2_{\rm s0} N^2_{\rm BV} \bar{\nu} B_{2\nu}}{g} \\
\nonumber
- l^2 c^2_{\rm s0} \bar{\nu}^2 (B_{2\nu} +  A_{1\nu}) &+& \bar{\nu}^3 B_{2\nu}(B_{1\nu}A_{2\nu} - A_{1\nu}B_{2\nu}) \\
\nonumber
2k^2 c^2_{\rm s0}\bar{\nu} B_{2\nu} + 2ikg \bar{\nu} B_{2\nu} &-& igl\bar{\nu} A_{2\nu} +kl\bar{\nu}c^2_{\rm s0}(B_{1\nu} + A_{2\nu})\Big]\\
\nonumber
{(i \omega)}\Big[\frac{N^2 \bar{\nu}^2 B^2_{2\nu}}{g} (il - ik) &+& \bar{\nu}^2l^2c^2_{\rm s0}(A_{1\nu}B_{2\nu} - A_{2\nu}B_{1\nu})\\
\nonumber
- lk\bar{\nu}^2 B_{2\nu} c^2_{\rm s0}(B_{1\nu} &+& A_{2\nu}) - k^2c^2_{\rm s0} \bar{\nu}^2 B^2_{2\nu} \\
\nonumber
+ l^2c^2_{\rm s0}N^2_{\rm BV} + l^2\bar{\nu}^2c^2_{\rm s0}A_{2\nu}B_{1\nu} &+& igB_{2\nu} \bar{\nu}^2 (A_{2\nu} l - B_{2\nu} k)\Big]\\
- \bar{\nu} B_{2\nu}l^2c^2_{\rm s0}N^2_{\rm BV} &=& 0
\end{eqnarray}
where $\bar{\nu} = {\eta}_{\rm c} T^{\frac{5}{2}}_{70}/{\rho}_0$, $k_{\rm R} = \sqrt{k^2 + l^2}$ and note that $-{(i \omega)}$ is the net growth/decay rate with the imaginary part implying oscillations. If we take $\bar{\nu} = 0$, we derive back the dispersion relation for linearly coupled g-modes and sound waves:
\begin{eqnarray}
\nonumber
{(i \omega)}^5 + {(i \omega)}^3 \Big[ c^2_{\rm s0} k^2_{\rm R} - ikg\Big(1+ \frac{c^2_{\rm s0} N^2_{\rm BV}}{g^2}\Big)\Big] + {(i \omega)} l^2c^2_{\rm s0}N^2_{\rm BV} = 0
\end{eqnarray}
This has two solutions (${\omega}^2 \approx \frac{l^2 N^2_{\rm BV}}{k^2_{\rm R}}$) for low-frequency pure oscillations which correspond to the g-modes. A pair of high-frequency pure oscillatory solutions (${\omega}^2 \approx c^2_{\rm s0} k^2_{\rm R}$) corresponds to the sound waves (note however, that the term $ikg$ is associated with attenuation of the wave and the acoustic cut-off frequency, or in other words the scale height $H = c^2_{\rm s0}/\gamma g$; in presence of g-modes this is further modified). If $k_{\rm R} c_{\rm s0} \sim N_{\rm BV}$, ${\omega}^2 = 0.5 c^2_{\rm s0} k^2_{\rm R} \pm 0.5 c^2_{\rm s0} k^2_{\rm R} \sqrt{1 - \frac{4 l^2 N^2_{\rm BV}}{c^2_{\rm s0} k^4_{\rm R}}} \sim 0.5 c^2_{\rm s0} k^2_{\rm R} \pm 0.5 c^2_{\rm s0} k^2_{\rm R} \sqrt{1 - \frac{4 l^2}{k^2_{\rm R}}}$. The last expression simply provides two sets of sound waves (${\omega}^2=c^2_{\rm s0}l^2$ and $c^2_{\rm s0}k^2$) when $l<<k \approx k_{\rm R}$. Also, note that $l \approx k_{\rm R}>>k$ implies that these modes are growing/decaying (negative discriminant). On the other hand, if $k_{\rm R} c_{\rm s0} \gtrsim N_{\rm BV}$, there is a solution : ${\omega}^2 = 0.5 c^2_{\rm s0} k^2_{\rm R} \pm 0.5 c^2_{\rm s0} k^2_{\rm R} \sqrt{1 - \frac{4 l^2 N^2_{\rm BV}}{c^2_{\rm s0} k^4_{\rm R}}}$ which gives two pairs of waves ${\omega}^2=c^2_{\rm s0} k^2_{\rm R} - \frac{l^2 N^2_{\rm BV}}{k^2_{\rm R}}$ and ${\omega}^2=\frac{l^2 N^2_{\rm BV}}{k^2_{\rm R}}$ ($l<<k \approx k_{\rm R}$) - modified sound waves and g-modes. With $l \approx k_{\rm R}>>k$, there are two pairs of sound waves modified by g-modes as well. Thus depending on wavenumbers $l$, $k$ and the Brunt-V\"ais\"al\"a frequency, there is reflection or growth or decay of waves. This linear physics is limited but it hints at how sound waves may alter in presence of other inhomogeneities in the medium. In some cases, sound waves are amplified while in others these decay. In reality, at this stage possibly sound waves become non-linear which is outskirts of core for a galaxy cluster. 

\section{With and without dynamic viscosity in the simulation}
\subsection{Compressible \& incompressible modes}
\label{sec:app1}
\begin{figure}
    \includegraphics[width=8cm]{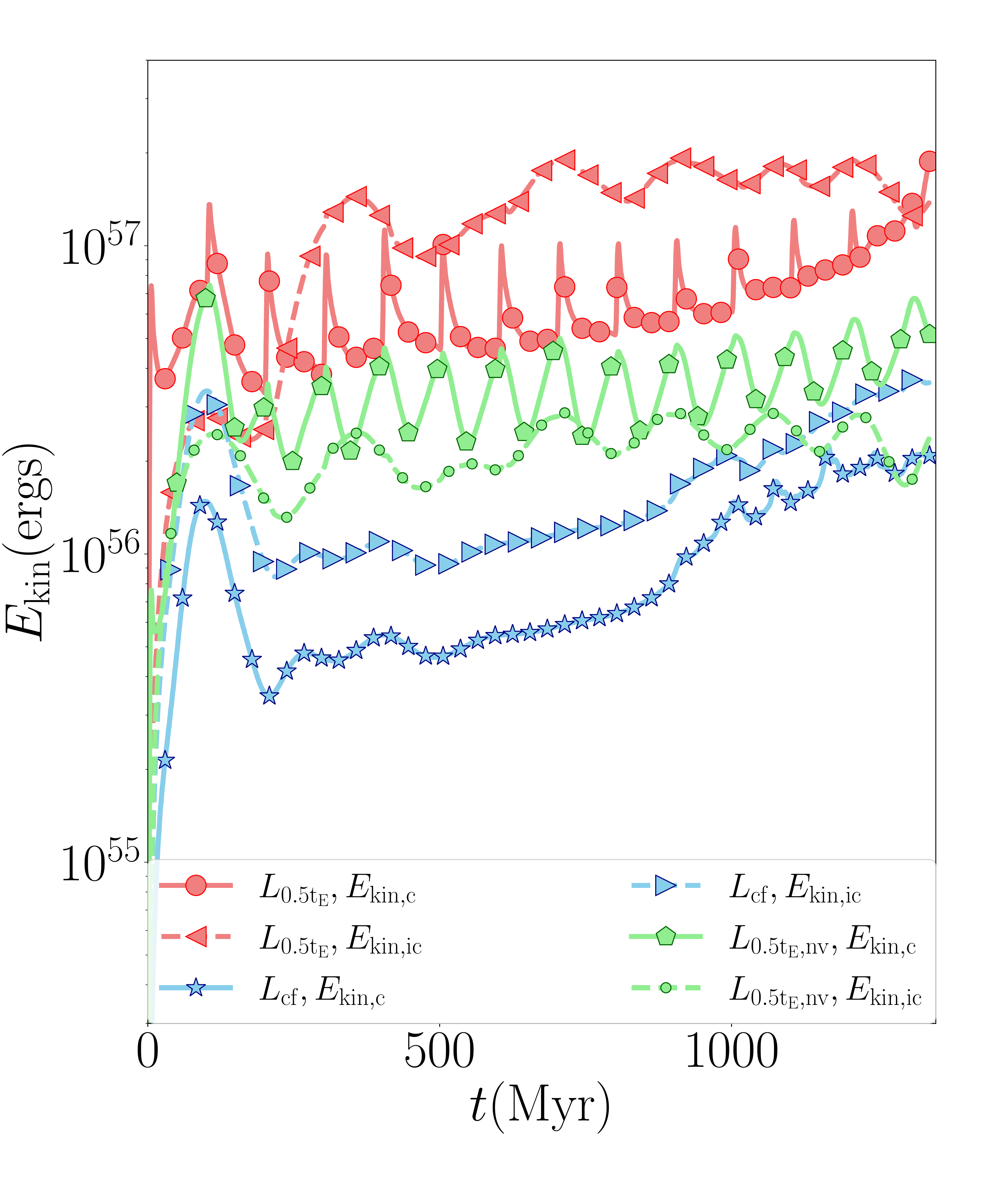}
    \centering
    \caption{An extension of Figure \ref{fig:KE}. Here we add two additional cases, cooling-flow and an ICM with no viscosity, $L_{\rm cf}$ and $L_{\rm 0.5t_E, nv}$ respectively. The former (blue) has only viscous heating, the latter (green) has only injected $E_{\rm inj}$. A combined heating (red) triggers significantly larger kinetic energies in the box. Note the gradual rise in the blue lines correspond to a cooling flow around $1$ Gyr (faster collapse than the other two cases).}
    \label{fig:KE_extend}
\end{figure}

We plot the compressible and incompressible kinetic energy for some additional cases in Figure \ref{fig:KE_extend} relative to discussions associated with Figure \ref{fig:KE}. The case of cooling flow, $L_{\rm cf}$ (also see Table \ref{tab:sims}) does not show waves since there is no injection of thermal energy into the medium. At very late times, there is a slight increase in the kinetic energy at the onset of cooling flow (note that there is no sink in our set-up and hence matter accumulates within an extremely small central volume around this time). It is interesting to note two characteristic differences between red and green lines: (i) the compressible features locally get attenuated (solid lines) with viscosity (red), and (ii) both compressible and incompressible velocities are raised at least by a factor of $2$ with viscosity (red). The extra diffusion in momentum and energy due to the viscous effects helps to add significantly larger kinetic modes in the box. This is possibly due to availability of larger amount of thermal energy when viscous dissipation works. The thermalization, in turn, generates additional swirling motions and waves. 

\subsection{Dissipation} 
\begin{figure}
    \includegraphics[width=8cm]{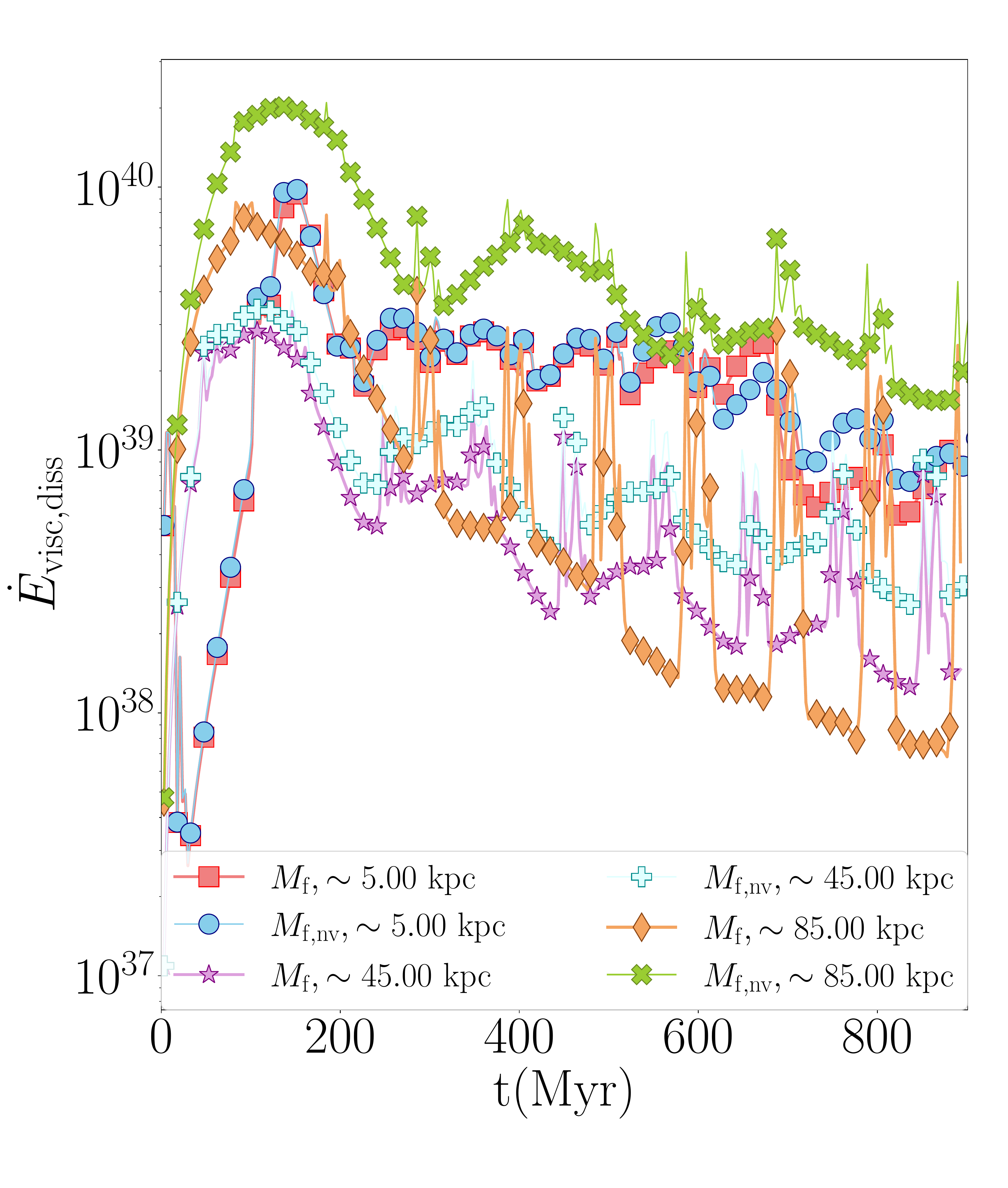}
    \centering
    \caption{Comparison of the dissipation rates at different radii with ($M_{\rm f}$) and without ($M_{\rm f, nv}$) adding viscous diffusion terms in the simulation.}
    \label{fig:diss_extend}
\end{figure}
Contrary to the result in last section in which we find that kinetic energy of the waves is enhanced in presence of a dynamic viscosity, here we discuss how dissipation of turbulence is overestimated without explicitly putting the viscous diffusion terms. Figure \ref{fig:diss_extend} shows $\dot{E}_{\rm visc, diss}$ for $M_{\rm f}$ and $M_{\rm f, nv}$. For latter, we essentially post-process dissipation rate assuming the same suppression factor ${\xi}_{\nu}=0.1$ that explicitly goes into the former while solving the hydrodynamic equations ($M_{\rm f, nv}$ does not use the RHS viscous terms in eqs \ref{eq:eq2}-\ref{eq:eq3}). At large radii (see green and orange lines/markers), the difference is enormous, particularly for non-acoustic dissipation. 

This exploration with $L_{\rm f, nv}$ (previous section) and $M_{\rm f, nv}$ implies that while dynamic viscous heating generates significantly higher acoustic or solenoidal modes in the medium, there is a stronger homogenization of velocities overall in presence of dynamic viscosity. It also highlights the role of sound waves in the outskirts very strongly. Naively, the green lines in Figure \ref{fig:KE_extend} and the lighter green line in Figure \ref{fig:diss_extend} would imply that sound waves escape leaving no significant impact. 

%
%
%
\bsp	
\label{lastpage}
\end{document}